\documentclass[12pt]{article}
\usepackage{amsmath}
\usepackage{graphicx}
\usepackage{longtable}
\usepackage{euscript}
\usepackage{color}
\usepackage{latexsym}
\usepackage{amsbsy}
\usepackage{amssymb}
\usepackage{soul}
\usepackage{booktabs}
\usepackage{tabularx}
\usepackage{float}
\usepackage{endfloat} % control plot position
\usepackage{times}
\usepackage{verbatim}
\usepackage{enumerate}
\usepackage{enumitem}
\usepackage{multirow}
\usepackage{setspace}
\usepackage{epsfig}
\usepackage{graphics}
\usepackage{amsthm}
\usepackage{amsfonts}
\usepackage{cancel}
\usepackage{algorithm}
\usepackage{algpseudocode}
\usepackage{multicol} 
\usepackage[style=apa,backend=biber,maxbibnames=20,minbibnames=1]{biblatex}
\renewbibmacro{in:}{}
\DeclareLanguageMapping{english}{english-apa}
\newtheorem{remark}{Remark}
\newtheorem{theorem}{Theorem}
\newtheorem{lemma}{Lemma}

\newcommand{\bm}[1]{\mbox{\boldmath{$#1$}}}
\newcommand{\be}{\begin{eqnarray}}
\newcommand{\ee}{\end{eqnarray}}
\newcommand{\ba}{\begin{eqnarray*}}
\newcommand{\ea}{\end{eqnarray*}}

\newcommand {\bfbeta} {\mbox{\boldmath $\beta$}}

\newcommand {\bftheta} {\mbox{\boldmath $\theta$}}

\newcommand {\bfI} {{\bf I}}

\newcommand {\bfY} {{\bf Y}}

\newcommand {\bfx} {{\bf x}}

\newcommand{\simiid}{\stackrel{\text{i.i.d}}{\sim}}

\usepackage{authblk}
\author[1,3]{Shixuan Wang\thanks{wang3sx@mail.uc.edu}}
\author[2]{Jing Zhang\thanks{zhangj8@miamioh.edu}}
\author[3]{Emily L. Kang\thanks{kangel@ucmail.uc.edu}}
\author[1,4]{Bin Zhang\thanks{Bin.Zhang@cchmc.org}}

\affil[1]{Division of Biostatistics and Epidemiology, Cincinnati Children's Hospital Medical Center, Cincinnati, Ohio, U.S.A.}
\affil[2]{Department of Statistics, Miami University, Oxford, Ohio, U.S.A.}
\affil[3]{Division of Statistics and Data Science, Department of Mathematical Sciences, University of Cincinnati, Cincinnati, Ohio, U.S.A.
\affil[4]{Department of Pediatrics, University of Cincinnati, College of Medicine, Cincinnati, Ohio, U.S.A.}\vspace{-5ex}}

\title{\Large\textbf{PPD-CPP: Pointwise predictive density calibrated-power prior in dynamically borrowing historical information}}

\date{}
\begin{document}
\maketitle

\begin{abstract}
\begin{spacing}{1}
    Incorporating historical or real-world data into analyses of treatment effects for rare diseases has become increasingly popular. A major challenge, however, lies in determining the appropriate degree of congruence between historical and current data. In this study, we devote ourselves to the capacity of historical data in replicating the current data, and propose a new congruence measure/estimand $p_{CM}$. $p_{CM}$ quantifies the heterogeneity between two datasets following the idea of the marginal posterior predictive $p$-value, and its asymptotic properties were derived. Building upon $p_{CM}$, we develop the pointwise predictive density calibrated-power prior (PPD-CPP) to dynamically leverage historical information. PPD-CPP achieves the borrowing consistency and allows modeling the power parameter either as a fixed scalar or case-specific quantity informed by covariates. Simulation studies were conducted to demonstrate the performance of these methods and the methodology was illustrated using the Mother's Gift study and \textit{Ceriodaphnia dubia} toxicity test.
\end{spacing} 
\vspace{+10ex}{\it Key Words: Bayesian predictive p-value,  Calibrated power prior, Congruence measure, Dynamic information borrowing} 
\end{abstract}

\newpage

\section{Introduction}
\begin{spacing}{1}
Borrowing information from historical or external studies for ongoing research has long been an important topic in clinical trials, and it has recently gained growing attention in other fields such as psychology \parencite{kaplan2023}, toxicology \parencite{zhang2022}, and political science \parencite{isakov2020}. In clinical settings such as pediatric drug development for rare diseases, conducting randomized controlled trials may not be feasible due to very small patient populations. In these cases, historical control data become critical for evaluating treatment efficacy. The recent U.S. Food and Drug Administration guidance on complex innovative designs \parencite{fda2020citd} recommends leveraging available control group data from phase II studies to help accelerate drug approval, while emphasizing the need for rigorous assessment of exchangeability between historical and current data (i.e., whether they follow the same distribution).

A state-of-the-art frequentist framework for incorporating historical information is the test-then-pool approach \parencite{viele2014,li2020}. In this method, an equivalence test is first conducted; if the null hypothesis is not rejected, the historical data are fully pooled with the current data. However, this “all-or-none” borrowing strategy is prone to power loss and inflated type I error when historical data are inappropriately discarded or pooled. Recent progress in frequentist approach have shifted toward selectively incorporating subsets of historical data into a joint analysis \parencite{gao2025}.

Viewing historical data as a form of prior information provides a natural connection to Bayesian approaches. One widely used method is the power prior (PP; \citeauthor{ibrahim2000}, \citeyear{ibrahim2000}), which incorporates historical information by raising the historical data likelihood to a power $\alpha \in [0,1]$ and combining it with the current data likelihood. The power parameter $\alpha$ can be treated as a random variable \parencite{chen2000,duan2006,ye2022}; however, specifying an appropriate prior for $\alpha$ remains an open problem. In particular, although noninformative or weakly informative priors are commonly used, they may excessively discount historical information even when the historical and current data are consistent \parencite{neuenschwander2009,pawel2023}. To address this, some recent work focused on developing relatively informative priors for $\alpha$ \parencite{shen2023,demartino2025}.

Alternatively, the power parameter can be treated as a fixed constant \parencite{ibrahim2015}. \textcite{lu2022} utilizes propensity scores (PS) and views patients with similar PS across trials as exchangeable, but PS adjustment only balances observed covariates and may not fully capture the true congruence between datasets. Calibrated power prior (CPP; \citeauthor{pan2017}, \citeyear{pan2017}) and elastic power prior (EPP; \citeauthor{jiang2023}, \citeyear{jiang2023}) view $\alpha$ as a function of a congruence measure. CPP and EPP directly quantify the distributional similarity between historical and current data but require additional use of historical data to tune the hyperparameter in such function. Moreover, the theoretical properties of congruence measures have received very limited investigation. Beyond the power prior framework, other forms of historical data informed prior have been proposed \parencite{hobbs2011,schmidli2014,jiang2023,alt2024}. The common characteristic among these approaches is to propose a discounting parameter that determines the exact level of historical borrowing. A proper congruence measure can help evaluate these parameters.

{In this work, we propose a new congruence measure and develop the pointwise predictive density-calibrated power prior (PPD-CPP). The proposed measure leverages the tail probability of marginal posterior predictive distributions to quantify how likely the current data can be replicated, given the historical data. Originally, posterior predictive $p-$values \parencite{meng1994,gelman1996} are designed to check model fit in Bayesian analysis and does not have a closed form expression. However, the proposed congruence measure assesses the distributional congruence between historical and current data and it has finite or asymptotic closed forms for data from normal populations. We study its theoretical properties when historical and current data are either congruent (i.e. from the same distribution) or incongruent (mean difference; variance ratio difference; covariate shift). We show that the proposed measure converges to distinct point masses, in contrast to the uniform distribution of frequentist $p$-values. This distinctive property enables more flexible borrowing of historical information. Finally, we develop PPD-CPP, which considers the power parameter as a function of the proposed congruence measure. PPD-CPP can be viewed as a generalization of CPP \parencite{pan2017} where the tuning process of hyperparameters is no longer data dependent. When covariates are available, the proposed measure can evaluate the pointwise exchangeability and PPD-CPP can thus assign an individualized power parameter to each historical data observation. }

The remainder of the paper is organized as follows. Section~\ref{sec: pp} introduces the proposed congruence measure and explains how its theoretical properties support the construction of PPD-CPP, both in the univariate outcome setting and the regression framework for normal endpoints. Section~\ref{sec: simu} presents extensive simulation studies comparing PPD-CPP with alternative methods. Section~\ref{sec: real_data} illustrates the application of PPD-CPP using a vaccine trial and a toxicology experiment. Finally, Section~\ref{sec: disc} concludes with a discussion of the main findings.

\end{spacing}

\section{Pointwise predictive density-calibrated power prior}\label{sec: pp}
\begin{spacing}{1}

In this section, we begin by introducing the power prior and its calibrated variants. Next, in Section~\ref{subsec: measure}, we present how the proposed congruence measure is constructed based on the posterior predictive $p$-value and describe its theoretical properties. In Section~\ref{subsec: parameter}, we develop PPD-CPP based on the proposed congruence measure for normal endpoints, both with and without covariates.

\subsection{Power prior (PP)}\label{subsec: pp}

Let $\mathbf{Y}^h=(y^h_1,\ldots,y^h_m)^\top$ and $\mathbf{Y}^c=(y^c_1,\ldots,y^c_n)^\top$ denote the historical and current data, respectively, where $m$ and $n$ denote the corresponding sample sizes. Assume $\{y^h_i\}_{i=1}^m \stackrel{\text{iid}}{\sim} f_{\boldsymbol{\theta}_h}$ and $\{y^c_i\}_{i=1}^n \stackrel{\text{iid}}{\sim} f_{\boldsymbol{\theta}_c}$, where $\boldsymbol{\theta}_h$ and $\boldsymbol{\theta}_c$ are the parameter vectors for distribution $f$ corresponding to $\mathbf{Y}^h$ and $\mathbf{Y}^c$, respectively. Both $\mathbf{Y}^h$ and $\mathbf{Y}^c$ can be viewed as responses from the control arm. Under the assumption of exchangeability between historical and current data (i.e., $\boldsymbol{\theta} = \boldsymbol{\theta}_h = \boldsymbol{\theta}_c$), PP \parencite{ibrahim2000} is defined as:
\begin{equation}\label{eq: pp}
\pi(\boldsymbol{\theta}\mid \mathbf{Y}^h) \propto L(\boldsymbol{\theta}\mid \mathbf{Y}^h)^\alpha \pi_0(\boldsymbol{\theta})
\end{equation}
where $\pi_0(\boldsymbol{\theta})$ is an initial prior that is usually noninformative, $L(\boldsymbol{\theta}\mid \mathbf{Y}^h) = \prod_{i=1}^m f_{\boldsymbol{\theta}}(y^h_i)$ is the likelihood function based on the historical data, and $\alpha\in (0,1)$ is the power parameter that determines the degree of confidence in historical borrowing.

The power parameter $\alpha$ is often specified through a congruence measure $S \in (0,\infty)$ and a monotone decreasing link function $g(S)$ that maps $S$ to $\alpha$ via calibration. The congruence measure $S$ is typically defined as a distance metric (e.g., the Kolmogorov–Smirnov statistic) that decreases as the level of agreement between $\mathbf{Y}^h$ and $\mathbf{Y}^c$ increases. CPP and elastic prior methods \parencite{jiang2023} assume $g(S)$ follows a two-parameter sigmoid form:
\begin{equation}\label{eq:sigmoid}
    \alpha = g(S) = \frac{1}{1+\exp(a+b\log(S))}
\end{equation}
where $a \in \mathbb{R}$ and $b > 0$.

{To determine $a$ and $b$, the key challenge lies in characterizing how $S$ behaves under congruence versus incongruence. \textcite{pan2017} explore $S$ using two thresholds: $\gamma^C$, the maximum acceptable difference in mean for a sample deemed congruent with $\bfY^h$, and $\gamma^{IC}$, the minimum tolerated difference for a sample deemed incongruent. These thresholds enable the generation of samples classified as either congruent or incongruent with $\bfY^h$, therefore allowing the distribution of $S$ to be numerically summarized under both scenarios. The parameters $a$ and $b$ are then solved from Equation~\eqref{eq:sigmoid}. In practice, \textcite{pan2017} and \textcite{jiang2023} rely on expert knowledge to define $\gamma^C$ and $\gamma^{IC}$, while \textcite{zhang2024} and \textcite{wang2024} adopt simulation-based methods. Despite these efforts, two main limitations remain: (i) $S$ has primarily been studied numerically, with limited theoretical development of the underlying congruence measures; and (ii) the calibration process depends heavily on the historical data, which risks allowing historical information to dominate inference.}

\subsection{Posterior predictive $p$-value as the congruence measure with a desired null nonuniformity in historical borrowing}\label{subsec: measure}

     The posterior predictive $p$-value \parencite{meng1994,gelman1996}, conditional on $\bfY^h$, is defined as:
$$
    p_B = \Pr\left( T(\bfY^{rep}) \geq  T(\bfY^h) \mid \bfY^h \right)
    $$
where $\bfY^{rep}=(y^{rep}_1,\ldots,y^{rep}_m)^\top$ denotes posterior predictive replicates and $T(\cdot)$ is a sample statistic (e.g., maximum, quantile). As emphasized by \textcite{gelman1995book}, $p_B$ evaluates the degree of systematic misfit between the observed data and the posterior predictive replicates. In contrast to the frequentist $p$-value, which is uniformly distributed under the null, $p_B$ has a nonuniform null distribution that tends to concentrate around $1/2$. Values of $p_B$ that deviate toward 0 or 1 indicate growing disagreement between $\bfY^h$ and the replicates. When the model adequately represents the data-generating mechanism of $\bfY^h$, the posterior predictive samples $\bfY^{rep}$ provide a meaningful forecast of $\bfY^h$ \parencite{gelman2007,gelman2013}.

Motivated by this perspective, when $\bfY^h$ and $\bfY^c$ are congruent (i.e., drawn from the same probabilistic distribution), the behavior of $p_B$ with $T(\bfY^c)$ is expected to follow a similar pattern. This leads us to define a congruence measure, $p_{CM}$, as the posterior predictive $p$-value comparing $\bfY^c$ to $\bfY^h$:
\begin{equation}
p_{CM} = \Pr\left( T(\bfY^{rep}) \geq T(\bfY^c) \mid \bfY^h \right).
\label{pcm_1}
\end{equation}
Here, the subscript $CM$ stands for ``Congruence Measure.'' Intuitively, $p_{CM}$ measures how well the historical data can replicate the current data. Note that $\bfY^{rep}$ has the same dimension as $\bfY^c$ when conditioning on $\bfY^h$. Ideally, $p_{CM}$ approaches $1/2$ when $\bfY^c$ and $\bfY^h$ are highly congruent, and shifts toward 0 or 1 as incongruence increases. However, even when $\bfY^h$ and $\bfY^c$ originate from the same data-generating process (e.g., two independent trials under an identical protocol), random variation causes $p_{CM}$ to follow a uniform distribution under the null (i.e., congruence). Our simulations (see Appendix~\ref{app:a5}) confirm this property. This uniformity complicates adaptive borrowing of information, as values of $p_{CM}$ near 0 or 1—expected only under incongruence—can arise by chance. For this reason, frequentist and Bayesian methods that rely directly on $p$-values for historical borrowing \parencite{liu2018,kwiatkowski2024} require additional adjustment to correct for this behavior.

To address this issue, we propose using the marginal posterior predictive $p$-value \parencite{gelman1995book} when comparing the current data $\bfY^c$ to the historical data $\bfY^h$. In this setting, the congruence measure is redefined as
\begin{equation}\label{pcm}
\begin{aligned}
p_{CM} &= \Pr\left\{ T(y^{rep}_i) \geq T(y^{c}_i) \mid \bfY^h \right\}, \;\;\; \forall i=1,\ldots,n,
\end{aligned}
\end{equation}
where $y^{rep}_i$ denotes the $i$th entry of the posterior predictive sample $\bfY^{rep}$ and $y^c_i$ denotes the $i$th entry of the current data $\bfY^c$. It is important to note that $p_{CM}$ in (\ref{pcm}) is not indexed by $i$ (e.g., $p_{CM,i}$). The reason is that, conditional on $\bfY^h$, the elements $\{y^c_i\}_{i=1:n}$ and $\{y^{rep}_i\}_{i=1:n}$ are independent and identically distributed (i.i.d.) respectively when covariates are not considered. In this case, each $p_{CM,i}$ takes the same value, so there is no need to distinguish them. However, the situation changes once covariates are introduced. When covariates are present, the elements of $\{y^c_i\}_{i=1:n}$ and $\{y^{rep}_i\}_{i=1:n}$ are no longer identically distributed, since each observation depends on its associated covariate values. In this case, the congruence measure must therefore be indexed by $i$, and we will reintroduce the notation $p_{CM,i}$ under the regression setting with covariates in Section~\ref{subsec: cov}.

    {For the congruence measure in (\ref{pcm}), certain choices of $T(x)$, such as quantiles, are no longer applicable. We therefore consider two alternatives: $T(x) = x$ and $T(x; \bfY^h) = p(x \mid \bfY^h)$, where $p(x\mid\bfY^h)=\int \text{L}( \bftheta\mid x)\pi(\bftheta\mid \bfY^h) d\bftheta$ denotes the marginal posterior predictive likelihood. The theoretical properties and simulation results of $p_{CM}$ under both choices are studied. In the main text, however, we focus on presenting results based on the latter choice.}

With $T(x;\mathbf{Y}^h)=p(x\mid \mathbf{Y}^h)$, we can reformulate (\ref{pcm}) as
\begin{equation}\label{pcm_mc} 
        \begin{aligned}
             p_{CM}& = \Pr\left\{ p(y^{rep}_i\mid \bfY^h) \geq  p(y^{c}_i\mid \bfY^h)  \right\} \\
             &= E_{(y^{rep}_i\mid\bfY^h),(y^c_i)} \left\{ {\bfI} \left[ p(y^{rep}_i\mid \bfY^h) \geq  p(y^{c}_i\mid \bfY^h) \right]  \right\} \\
             &= \int\int {\bfI} \left[ p(y^{rep}_i\mid \bfY^h) \geq  p(y^{c}_i\mid \bfY^h) \right] p(y^{rep}_i\mid \bfY^h) f_{\bftheta_c}(y^c_i)\;dy^{rep}_i dy^c_i\\
        \end{aligned}
    \end{equation}
where $\mathbf{I}[\cdot]$ denotes the indicator function, and we express $p_{CM}$ as the expectation of the binary random variable
$$
W_i = \mathbf{I}\!\left[p(y^{rep}_i\mid \mathbf{Y}^h) \geq p(y^{c}_i\mid \mathbf{Y}^h)\right], \quad i=1,\ldots,n,
$$
with $W_i \overset{iid}{\sim} \text{Bern}(p_{CM})$. The i.i.d. structure arises because given $\bfY^h$, both $\{y^c_i\}_{i=1}^n$ and $\{y^{rep}_i\}_{i=1}^n$ are i.i.d., respectively. It should be noted that $p_{CM}$ is an estimand that depends on $\boldsymbol{\theta}_c$ and that a closed-form expression is generally unavailable. {For normal endpoints, however, we derive a closed form for $p_{CM}$ and study its theoretic properties.}

    \begin{lemma}\label{lmma} 
        Let $\{y^h_i\}_{i=1:m}\simiid N(\mu_h,\sigma^2_h)$ and $\{y^c_i\}_{i=1:n} \simiid N(\mu_c,\sigma^2_c)$. $\sigma^2_h$ and $\sigma^2_c$ are known. Let the test statistic be the marginal posterior predictive likelihood, with $\pi_0(\mu_h)\propto 1$,
        \begin{equation*}
            p_{CM}=Pr \left\{ \begin{pmatrix}U\\ V\end{pmatrix} \geq \begin{pmatrix}0\\ 0\end{pmatrix} \right\} +Pr \left\{ \begin{pmatrix}U\\ V\end{pmatrix} \leq \begin{pmatrix}0\\ 0\end{pmatrix} \right\}
        \end{equation*}
        where $\begin{pmatrix}U\\ V\end{pmatrix} =\begin{pmatrix}y^c_i+y^{rep}_i-2\Bar{y}^h\\ y^c_i-y^{rep}_i\end{pmatrix}\sim MVN(\begin{pmatrix}\mu_c-\Bar{y}^h\\ \mu_c-\Bar{y}^h\end{pmatrix},\begin{bmatrix} \sigma^2_c+\frac{m+1}{m}\sigma^2_h &\sigma^2_c-\frac{m+1}{m}\sigma^2_h\\\sigma^2_c-\frac{m+1}{m}\sigma^2_h & \sigma^2_c+\frac{m+1}{m}\sigma^2_h\end{bmatrix})$. When $\sigma^2_h$ and $\sigma^2_c$ are unknown, with $\pi_0(\mu_h,\sigma^2_h)\propto \frac{1}{\sigma^2_h}$, $\begin{pmatrix}U\\ V\end{pmatrix} \sim MVN(\begin{pmatrix}\mu_c-\mu_h\\ \mu_c-\mu_h\end{pmatrix},\begin{bmatrix} \sigma^2_c+\sigma^2_h &\sigma^2_c-\sigma^2_h\\\sigma^2_c-\sigma^2_h & \sigma^2_c+\sigma^2_h\end{bmatrix})$ asymptotically.

        \begin{proof}
            See Appendix~\ref{app:a2}
        \end{proof}
    \end{lemma}
    Lemma \ref{lmma}. provides a special case for normal endpoints with known variance, where $y^{rep}_i\mid \bfY^h$ is normally distributed and thus $p_{CM}$ can be explicitly expressed with finite samples. In practice, we estimate $p_{CM}$ by the sample mean for historical/current data respectively. When the variances were unknown, $y^{rep}_i\mid \bfY^h$ is $t$ distributed and therefore making the density function of $U$ and $V$ analytically intractable. Nevertheless, an asymptotic closed form of $p_{CM}$ in Lemma \ref{lmma}. is still available for practical use. Note that even though we focus on normal responses, the idea behind $p_{CM}$ can be generalized to count, dichotomy too. To address the intractability issue arising either from the posterior predictive distribution of $y^{rep}_i\mid \bfY^h$ or the joint distribution of vector $(U, V)^\top$, we recommend using a Monte Carlo method to approximate $p_{CM}$ as 
    \begin{equation}\label{pcm_compu}
        \frac{1}{nR}\sum_{r=1}^{R}\sum_{i=1}^{n} {\bfI} \left[ p(y^{rep}_{i(r)}\mid \bftheta_{h(r)}) \geq  p(y^{c}_i\mid \bftheta_{h(r)})\right]
    \end{equation}
    where $R$ represents the number of markov chain monte carlo (MCMC) iterations. $\bftheta_{h(r)}$ denotes the posterior realization of $\bftheta_h$ at $r^{th}$ MCMC iteration conditional on $\bfY^h$. For the normal case with unknown variance, $\bftheta_h=(\mu_h,\sigma^2_h)^\top$. Approximating $p_{CM}$ via (3) also alleviates the concern of using the asymptotic form of $p_{CM}$ as the sample size from historical/current data will never reach infinity in practice.
    \begin{theorem}\label{keythm} 
    Let $\{y^h_i\}_{i=1:m}\simiid N(\mu_h,\sigma^2_h)$ and $\{y^c_i\}_{i=1:n} \simiid N(\mu_c,\sigma^2_c)$ be independent. Let test statistic be the marginal posterior predictive likelihood. Assume $\pi_0(\mu_h)\propto 1$ when $\sigma^2_h$ is known and  $\pi_0(\mu_h,\sigma^2_h)\propto \frac{1}{\sigma^2_h}$ when $\sigma^2_h$ are unknown. For known variance case, $\bftheta_h=\mu_h$ and $\bftheta_c=\mu_c$; for unknown variance case, $\bftheta_h=(\mu_h,\sigma^2_h)^\top$ and $\bftheta_c=(\mu_c,\sigma^2_c)^\top$. When historical data and current data are congruent (i.e. $\bftheta_h=\bftheta_c$), regardless of known or unknown variance, 
        $$
        p_{CM} = \frac{1}{2}
        $$
    as $m\rightarrow \infty$. When data present growing incongruence (i.e. $\lvert\mu_c-\mu_h\rvert \rightarrow\infty$ or $\log\left( \frac{\sigma_c^2}{\sigma_h^2} \right) \to \infty$),
    $$
    p_{CM} = 1
    $$
    as $m\rightarrow \infty$.
        \begin{proof}
            See Appendix~\ref{app:a2}
        \end{proof}
    \end{theorem}
    \begin{remark}
         The incongruence arising from $\log\left( \frac{\sigma_c^2}{\sigma_h^2} \right) \to \infty$ represents the current data becomes increasingly uninformative and uncertain relative to the historical data. Additionally, we show $p_{CM} = 0$ as $m\rightarrow \infty$ when $\log\left( \frac{\sigma_h^2}{\sigma_c^2} \right) \to \infty$.  However, this case of incongruence is trivial, as the historical data becomes uninformative for the current data and therefore the need for historical borrowing is reduced.
    \end{remark}
    \begin{remark}
        Flipping the sign in (\ref{pcm_mc}) (i.e. $p(y^{rep}_i\mid \bfY^h) \leq  p(y^{c}_i\mid \bfY^h) $) does not affect the theoretical properties under congruence. However, we will obtain $p_{CM}=0$ as  $\lvert\mu_c-\mu_h\rvert \rightarrow\infty$ or $\log\left( \frac{\sigma_c^2}{\sigma_h^2} \right) \to \infty$, and $p_{CM}=1$ as $\log\left( \frac{\sigma_h^2}{\sigma_c^2} \right) \to \infty$ under asymptotics. The sign change will not alter the fact $\lvert p_{CM}-\frac{1}{2}\rvert \in (0,\frac{1}{2})$.
    \end{remark}
    Theorem \ref{keythm}. provides theoretical justification for why $p_{CM}$ exhibits nonuniformity concentrated around $1/2$ when data are congruent but converging to point masses at $0$ or $1$ when data are incongruent under asymptotics. Therefore, calibration process built upon $p_{CM}$ in (\ref{pcm_mc}) is free of any uniformity concerns under congruence. This result is established using the posterior predictive likelihood as $T(x)$ for normal endpoints and serves as the core of PPD-CPP. A similar theorem with $T(x)=x$ is also derived in the Appendix~\ref{app:a3}.

    \subsection{Power parameter $\alpha$ as a known scalar} \label{subsec: parameter}
    Let $W=\sum_{i=1}^{n} W_i=\sum_{i=1}^{n}{\bf 1} \left[ p(y^{rep}_i\mid \bfY^h) \geq  p(y^{c}_i\mid \bfY^h) \right]$ be the number of posterior predictive samples which are more likely to observe than the current data. Since $W_i\simiid Bern(p_{CM})$, for finite samples, we assume $W\sim \text{Binom}(n,p_{CM})$ and therefore a natural estimator of $p_{CM}$ is $W/n$. When $\bfY^h$ and $\bfY^c$ are congruent, the inherent nonuniformity in Theorem \ref{keythm}. allows us to assume $W^C\sim \text{Binom}(n,p_1=1/2)$. When $\bfY^h$ and $\bfY^c$ are completely incongruent, the ``direction" of likelihood function suggests $W^{IC} \sim \text{Binom}(n,p_2=1)$. In practice, we let the congruence measure be $S=\lvert p_{CM}-\frac{1}{2}\rvert$ in (\ref{eq:sigmoid}), where $\lvert p_{CM}-\frac{1}{2}\rvert \in (0,1/2)$ ensures a strict monotone decreasing relationship between $\alpha$ and itself. Note that assuming $W^{IC} \sim \text{Binom}(n,p_2=0)$ leads to the same interpretation since it does not vary the range of $\lvert p_{CM}-\frac{1}{2}\rvert$. Based on the above formulation, we manage to derive the closed-form distribution of $S$ under both congruence and incongruence scenario, and they are $\lvert \frac{W^C}{n}-\frac{1}{2}\rvert$ and $\lvert \frac{W^{IC}}{n}-\frac{1}{2}\rvert$ respectively. These distributions depend only on current sample size $n$ and are independent of any observed historical/current information. Therefore, we propose calibrating $a$ and $b$ data-independently as 
    \begin{equation}\label{cali_1}
        \begin{cases}
            \alpha^C = \frac{1}{1+\exp\left[a+b\log\left\{\lvert \mathbb{E} \left(\frac{W^C}{n}\right )-\frac{1}{2}\rvert \right\}\right]}\\
            \alpha^{IC} = \frac{1}{1+\exp\left[a+b\log\left\{\lvert \mathbb{E}\left(\frac{W^{IC}}{n}\right )-\frac{1}{2}\rvert \right\}\right]}
        \end{cases}       
    \end{equation}
    where $\mathbb{E}(\cdot)$ denotes the expectation. We use, for example, $\lvert \mathbb{E} \left(\frac{W^{C}}{n}\right )-\frac{1}{2}\rvert$ instead of $\mathbb{E} \left(\lvert \frac{W^C}{n}-\frac{1}{2} \rvert \right )$ because Jensen's inequality results in a more accurate estimate. In practice, other choices such as median (calibrated power prior, \textcite{pan2017}) or quantile (elastic prior, \textcite{jiang2023}) are also feasible. Let the $95\%$ confidence interval of $p_1,p_2$ denote as $(L_1,U_1)$ and $(L_2,U_2)$ respectively. To borrow almost congruent information (i.e. maximize the power) and discard nearly incongruent data (i.e. control type-I error rate), we further propose calibrating $a$ and $b$ by 
    \begin{equation}\label{cali_2}
        \begin{cases}
            \alpha^C = \frac{1}{1+\exp\left[a+b\log\left\{\lvert \mathbb{E} \left(\frac{W^C}{n}\right )+k_1-\frac{1}{2}\rvert \right\}\right]}\\
            \alpha^{IC} = \frac{1}{1+\exp\left[a+b\log\left\{\lvert \mathbb{E}\left(\frac{W^{IC}}{n}\right )-k_2-\frac{1}{2}\rvert \right\}\right]}
        \end{cases}       
    \end{equation}
    where $k_1 = {\max (\lvert L_1-1/2 \rvert,\lvert U_1-1/2 \rvert)}/\tau $ and $k_2 = {\max (\lvert L_2-1/2 \rvert,\lvert U_2-1/2 \rvert)}/{\tau}$.  $\tau$ is a confidence parameter reflecting our belief of the calibration process. $k_1$ and $k_2$ can be viewed as a form of ``standard error" for $p_{CM}$, capturing its uncertainty and depending solely on the current sample size $n$. 
    
\begin{figure}
            \centerline{
            \includegraphics[scale=0.12]{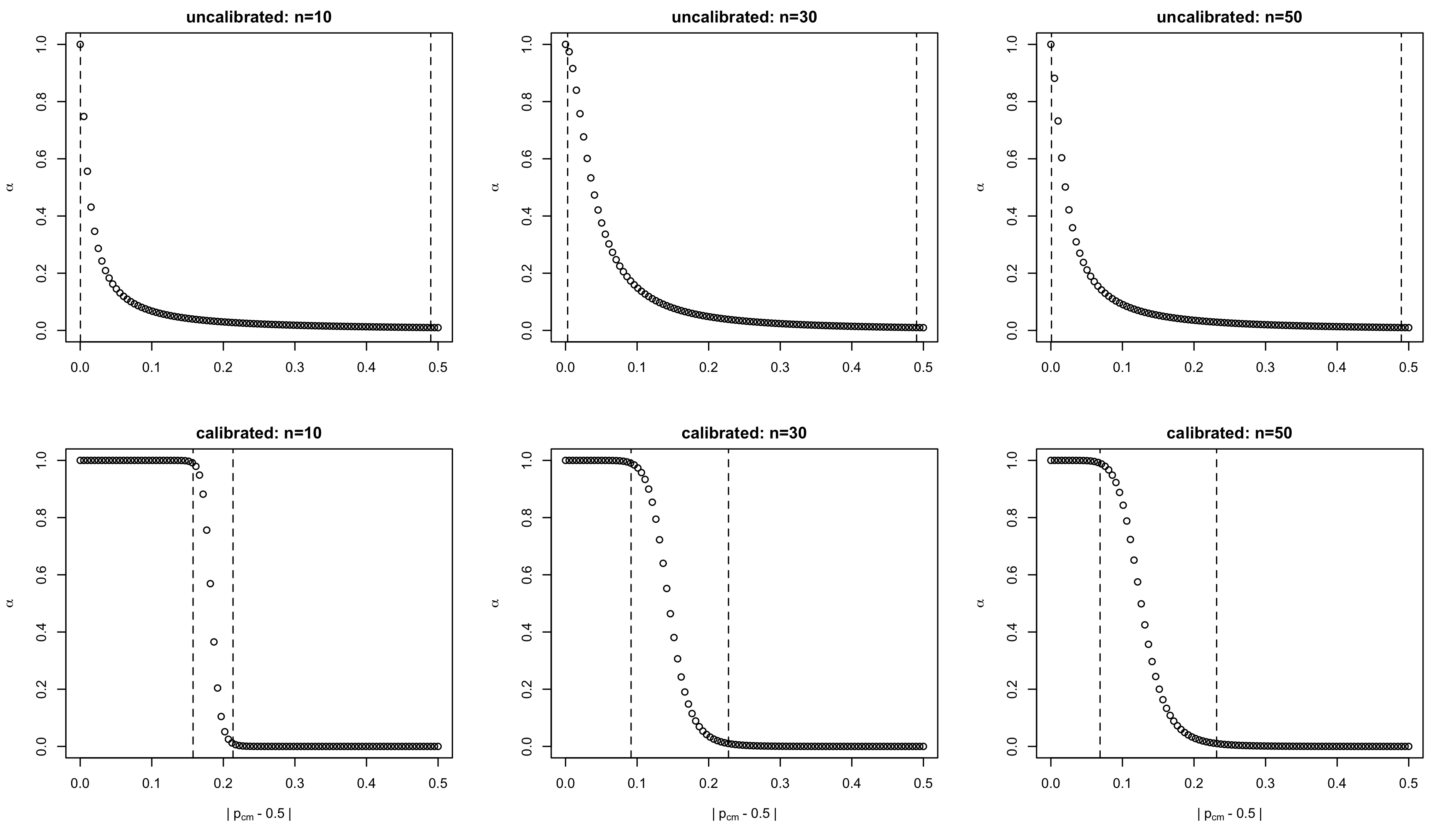}
            }
            \caption{Power parameter as a function of $\lvert p_{CM}-\frac{1}{2}\rvert$. The ``uncalibrated" method is based on (4) while the proposed ``calibrated" one is based on (5). The dashed curves on the left represent the cutoff when the power parameter $\alpha$ is greater than $\alpha^C=0.99$. The dashed curves on the right represent the cutoff when the power parameter $\alpha$ is less than $\alpha^{IC}=0.01$. The position of dashed curves depends on $k_1$ and $k_2$.}
            \label{fig:1}
\end{figure}

    Figure~\ref{fig:1} illustrates how $n$ influences the proposed calibration process. When $n$ is small, both $k_1$ and $k_2$ are relatively large, leading to a higher probability of fully borrowing historical information while also maintaining greater sensitivity to incongruence. As $n$ increases, the relationship between $\alpha$ and $\lvert p_{CM}-\frac{1}{2}\rvert$ transitions from a stepwise function to a more elastic one by introducing grey areas that reflect partial historical borrowing. This shift can be interpreted as the current data becomes increasingly informative in assessing the true congruence between historical data and itself. 
    The existence of $k_1$ and $k_2$ provides flexible control over the historical borrowing pattern. For example, we can simply let $k_1=0$ if a conservative flavor was desired. When both $k_1$ and $k_2$ are $0$, the calibration of (\ref{cali_2}) reduces to the form of (\ref{cali_1}), and it tends to incorporate less historical information as the degree of congruence grows. A particularly appealing feature of the proposed calibration procedure is that the tuning parameters $a$ and $b$ are calibrated independently from either $\bfY^h$ or $\bfY^c$. This is because the distribution of $S$ can be fully specified under both congruent and incongruent scenarios, with assumption $W\sim \text{Binom}(n,p_{CM})$. This unique feature prevents overfitting or overreliance on historical information, and therefore ensuring the historical borrowing is determined in a robust and pre-specified manner. The following theorem provides a theoretical guarantee of borrowing consistency using the proposed calibration method.
    \begin{theorem}{\textbf{(Borrowing Consistency)}}\label{2ndthm} 
        When data are congruent (i.e. $\bftheta_h=\bftheta_c$), the proposed PPD-CPP achieves full information borrowing with $\alpha=1$; when data are becoming more incongruent, PPD-CPP tends to completely disregards historical information with $\alpha$ converging to $0$.
        \begin{proof}
            Appendix~\ref{app:a4}
        \end{proof}
    \end{theorem}

    \subsection{Historical borrowing in regression}\label{subsec: cov}
    In this section, we discuss the application of the proposed PPD-CPP when covariates are present from two perspectives: 1. Applying a single power parameter to the historical data as a whole; 2. Assigning a unique power parameter to each historical observation. 
    \begin{lemma}\label{lmma2} 
        Let $y_i^h \sim N\left( {\mathbf{x}_i^{h}}^\top \bm{\beta}_h, \sigma_h^2 \right),i=1,\hdots,m$ and $y_i^c \sim N\left( {\mathbf{x}_i^{c}}^\top \bm{\beta}_c, \sigma_c^2 \right), i=1,\hdots,n$ where $\mathbf{x}_i^{h},\mathbf{x}_i^{c}$ are the $p\times 1$ covariate vectors, $\bm{\beta}_h,\bm{\beta}_c$ are the $p\times 1$ vectors of regression parameters, $\sigma_h^2$ and $\sigma_c^2$ are unknown. $p$ is the number of regression coefficients. Let the test statistic be the marginal posterior predictive likelihood, with $\pi(\bm{\beta}_h, \sigma_h^2) \sim (\sigma_h^2)^{-\frac{p+2}{2}}$, the asymptotic closed form of the pointwise $p_{CM}$, $p_{CM,i}$, is
        \begin{equation*}
            p_{CM,i}=Pr \left\{ \begin{pmatrix}U_i\\ V_i\end{pmatrix} \geq \begin{pmatrix}0\\ 0\end{pmatrix} \right\} +Pr \left\{ \begin{pmatrix}U_i\\ V_i\end{pmatrix} \leq \begin{pmatrix}0\\ 0\end{pmatrix} \right\}
        \end{equation*}
       with  $
     \begin{pmatrix}
        U_i \\ V_i
    \end{pmatrix}\sim \mathcal{MVN} \left(
        \begin{pmatrix}
        {\mathbf{x}_i^{c}}^\top (\bm{\beta}_c - \bm{\beta}_h) \\
        {\mathbf{x}_i^{c}}^\top (\bm{\beta}_c - \bm{\beta}_h)
        \end{pmatrix},
        \begin{bmatrix}
        \sigma_c^2 + \sigma_h^2 H_i & \sigma_c^2 - \sigma_h^2 H_i \\
        \sigma_c^2 - \sigma_h^2 H_i & \sigma_c^2 + \sigma_h^2 H_i
        \end{bmatrix}
        \right)
    $
    where $H_i = 1 + {\mathbf{x}_i^{c}}^\top ({X^{h}}^\top X^h)^{-1} \mathbf{x}_i^c$ and $i=1,\hdots,n$. $X^h$ is the $m\times p$ design matrix for historical data.

        \begin{proof}
            Appendix~\ref{app:a2}
        \end{proof}
    \end{lemma}
    Lemma \ref{lmma2} provides an asymptotic closed-form expression for the pointwise $p_{CM}$ when borrowing historical information with covariates. Unlike Lemma \ref{lmma}, we demonstrate $p_{CM}$ of being pointwise because $y^h_i$'s or $y^{rep}_i$'s or $y^c_i$'s are no longer i.i.d when covariates are present. We therefore aggregate $p_{CM,i}$ and take $p_{CM}=\sum_{i=1}^{n}p_{CM,i}/n$ in practice. It is also worth noting that the pointwise $p_{CM}$ accounts for extrapolation risk through the term $ {\mathbf{x}_i^{c}}^\top ({X^{h}}^\top X^h)^{-1} \mathbf{x}_i^c$, a form of leverage statistic \parencite{chatterjee1986} which measures the pointwise deviation of current data on the historical input space. Regardless of the asymptotics, $\mathbf{x}_i^c$ may not necessarily be an interior point of the set $\{ \mathbf{x}_i^h\}_{i=1,\hdots,m}$ and therefore the range of the leverage statistic will not be upper bounded by $1$. This observation introduces another type of incongruence with respect to covariate shifts. In the Appendix~\ref{app:a2}, we examine different types of incongruence and demonstrate the same conclusion as Theorem \ref{keythm} and Theorem \ref{2ndthm} when considering covariates. Therefore, the calibration using either (\ref{cali_1}) or (\ref{cali_2}) remains valid even with covariates.

   To allow individualized weighting on historical observations, we harvest the feature of pointwise $p_{CM}$ as Lemma \ref{lmma2} and build PPD-CPP upon the goodness of replicating the historical data conditional on the current data. The specific steps are given in Algorithm~\ref{algo: ppd-cpp}.
   
   \begin{algorithm}[H]
    \caption{PPD-CPP: Assigning each historical observation a unique power parameter} \label{algo: ppd-cpp}
    %Historical data $\{y^h_i,\bfx^h_i \}_{i=1,\hdots,m}$, current data $\{y^c_i,\bfx^c_i \}_{i=1,\hdots,n}$
    \begin{algorithmic}[1]
    \State $\textbf{Data:}$ Historical data is $D^h=(\bfY^h,X_h)$ where $\bfY^h=(y^h_1,\hdots,y^h_m)$ and $X_h=( \bfx^h_1 \hdots \bfx^h_p)^\top$;  current data is $D^c=(\bfY^c,X_c)$ where $\bfY^c=(y^c_1,\hdots,y^c_m)$ and $X_c=( \bfx^c_1 \hdots \bfx^c_p)^\top$;
    \State $\textbf{Estimates:}$ $\hat{\bm{\beta}}_h=(X_h^\top X_h)^{-1}X_h^\top\bfY^h$; $\hat{\bm{\beta}}_c=(X_c^\top X_c)^{-1}X_c^\top\bfY^c$; $\hat{\sigma}^2_h=\frac{{\bfY^{h}}^\top (I_p - P^h) \bfY^h}{m-p}$ where $P^h=X^h ({X^{h}}^\top X^h)^{-1} {X^{h}}^\top$; $\hat{\sigma}^2_c=\frac{{\bfY^{c}}^\top (I_p - P^c) \bfY^c}{n-p}$ where $P^c=X^c ({X^{c}}^\top X^c)^{-1} {X^{c}}^\top$; $H_i = 1 + {\mathbf{x}_i^{h}}^\top ({X^{c}}^\top X^c)^{-1} \mathbf{x}_i^h$; $a,b$ as calibrated in (4);
    
    \For{$i = 1$ to $m$}
            \State $
            \hat{p}_{CM,i}=Pr \left\{ \begin{pmatrix}U_i\\ V_i\end{pmatrix} \geq \begin{pmatrix}0\\ 0\end{pmatrix} \right\} +Pr \left\{ \begin{pmatrix}U_i\\ V_i\end{pmatrix} \leq \begin{pmatrix}0\\ 0\end{pmatrix} \right\}
        $ where \\ \;\;\;\;\;\;\;\;\;\;\;\; $
     \begin{pmatrix}
        U_i \\ V_i
    \end{pmatrix}\sim \mathcal{MVN} \left(
        \begin{pmatrix}
        {\mathbf{x}_i^{h}}^\top (\hat{\bm{\beta}}_h - \hat{\bm{\beta}}_c) \\
        {\mathbf{x}_i^{h}}^\top (\hat{\bm{\beta}}_h - \hat{\bm{\beta}}_c)
        \end{pmatrix},
        \begin{bmatrix}
        \hat{\sigma}_h^2 + \hat{\sigma}_c^2 H_i & \hat{\sigma}_h^2 - \hat{\sigma}_c^2 H_i \\
        \hat{\sigma}_h^2 - \hat{\sigma}_c^2 H_i & \hat{\sigma_h}^2 + \hat{\sigma}_c^2 H_i
        \end{bmatrix}
        \right);
    $
    \State$\alpha_i = \frac{1}{1+\exp\left\{a+b\log(\lvert \{\hat{p}_{CM,i}-\frac{1}{2}\rvert )\right\}}$;
        \EndFor
    \end{algorithmic}
    \end{algorithm}

   \end{spacing}
    
\section{Simulation studies}\label{sec: simu}   
\begin{spacing}{1}
    In this section, we conduct simulations to examine the borrowing pattern and efficacy of the proposed PPD-CPP focusing on normal endpoints with or without covariates. 
    %we first investigate how current sample size and incongruence level would jointly affect the historical borrowing using the proposed calibration process.
    \subsection{Simulation setup}
    
    The simulation scenarios for normal endpoints are presented in the followings:  
        \begin{enumerate}
            \item We generate current data from $\{y^c_i\}_{i=1:n} \simiid \mathcal{N}(\mu_c,\sigma^2_c)$ and historical data from $\{y^h_i\}_{i=1:m}\simiid \mathcal{N}(\mu_h,\sigma^2_h)$ where the absolute mean difference $\lvert \mu_c-\mu_h \rvert$ takes values in $(-4,4)$ and we fix $\mu_c=20$. $\sigma^2_h=\sigma^2_c= 0.5^2$ are assumed known, and $n=m\in \{10,50\}$ but $n\leq m$.

            \item Same simulation setups are applied as above but we assume $\sigma^2_h=\sigma^2_c= 0.5^2$ are unknown.

            \item We generate current data from $y^c_i \sim \mathcal{N}(\beta^c_0+\beta^c_1x^c_{1i}+\beta^c_2x^c_{2i},\sigma^2_c)$ where $\sigma^2_c=0.5^2$ and $i=1,\hdots,n$. We generate historical data from $y^h_i \sim \mathcal{N}(\beta^h_0+\beta^h_1x^h_{1i}+\beta^h_2x^h_{2i},\sigma^2_h)$ where $\sigma^2_h=0.5^2$ and $i=1,\hdots,m$. $\sigma^2_h$ and $\sigma^2_c$ are unknown. We let $n=m=50$, $x^c_{1i},x^h_{1i}\sim Bern(0.5)$, and $x^c_{2i},x^h_{2i}\sim DU(40,70)$. We fix $\beta^c_0=50,\beta^c_1=8,\beta^c_2=0.5$ while varying the regression coefficients for historical data.
      
        \end{enumerate}

    $\{y^c_i\}_{i=1:n}$ and $\{y^h_i\}_{i=1:m}$ can be viewed as current/historical trial data from the same group (i.e. control group). We denote the proposed method as PPD-CPP-sim-lik, PPD-CPP-sim-obs, PPD-CPP-thm-lik, PPD-CPP-thm-obs, PPD-CPP-pw-obs, PPD-CPP-pw-obs. ``obs" and ``lik" refer to the test statistics $T(x)=x$ and $T(x;\bfY^h)=p(x\mid \bfY^h)$ respectively. The term ``sim" means $p_{CM}$ is estimated through \eqref{pcm_compu}, while ``thm" represents the method derived in Lemma 1 and 2. ``pw" denotes the use of pointwise power parameters assigned to individual historical observations as described in Algorithm 1, and we only consider this approach in the presence of covariates. For comparison, we include complete pooling method, fitting solely based on current data, CPP with Kolmogorov–Smirnov statistic, and EPP with scaled $T$ statistic as the competitor method. To calibrate EPP and CPP, we follow the 0.8-1.25 rule and let $\gamma^{C}=0$ and $\gamma^{IC}=0.223$.

    The Bayesian model is fitted using $\textit{rstan}$ package \parencite{rstan2025} with $6,500$ MCMC iterations and $1,500$ burn-in iterations. Since exchangeability assumption, we let the prior $\pi_0(\bftheta)\propto 1$ for known variance and $\pi_0(\bftheta)\propto 1/\sigma^2$ for unknown variance. To examine the pattern of historical borrowing, we report the average power parameter, the probability of complete borrowing (defined as the proportion of $\alpha>\alpha^{C}$), and the probability of entirely discarding historical information (defined as the proportion of $\alpha<\alpha^{IC}$), over $500$ power parameter estimates. We choose $\alpha^{C}=0.99$ and $\alpha^{IC}=0.01$. The model performance is evaluated by the average point estimation bias, average posterior standard deviation, coverage probability, and average interval width of the $95\%$ credible interval based on $500$ replicates for each simulation setup. 

    \subsection{Results}
\begin{figure}
            \centerline{
            \includegraphics[scale=0.1]{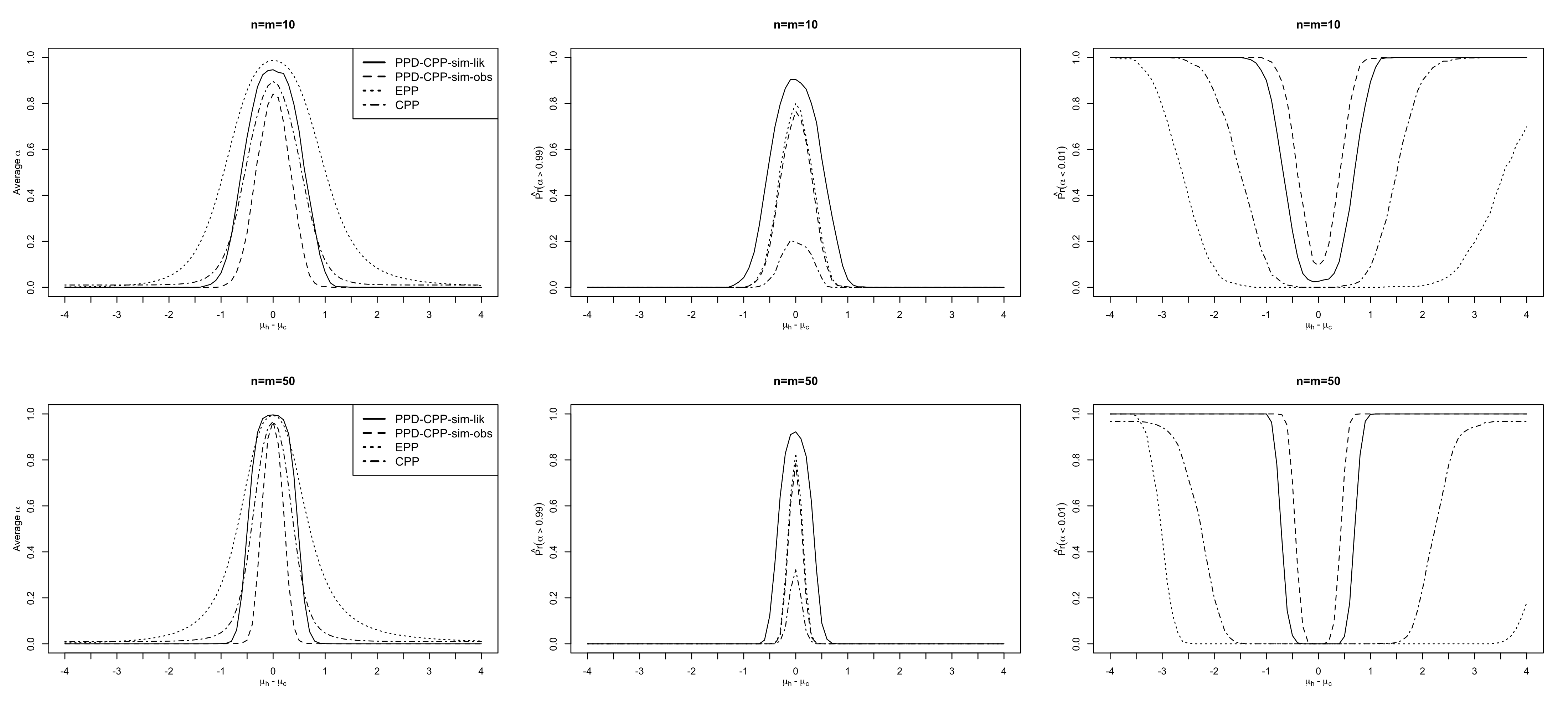}
            }
            \caption{Historical borrowing behavior with the change of mean difference for normal endpoints (no covariates), where $\sigma^2_h$ and $\sigma^2_c$ are assumed known. The value on the curve is computed using $500$ power parameters.}
            \label{fig:2}
\end{figure}

\begin{figure}
            \centerline{
            \includegraphics[scale=0.1]{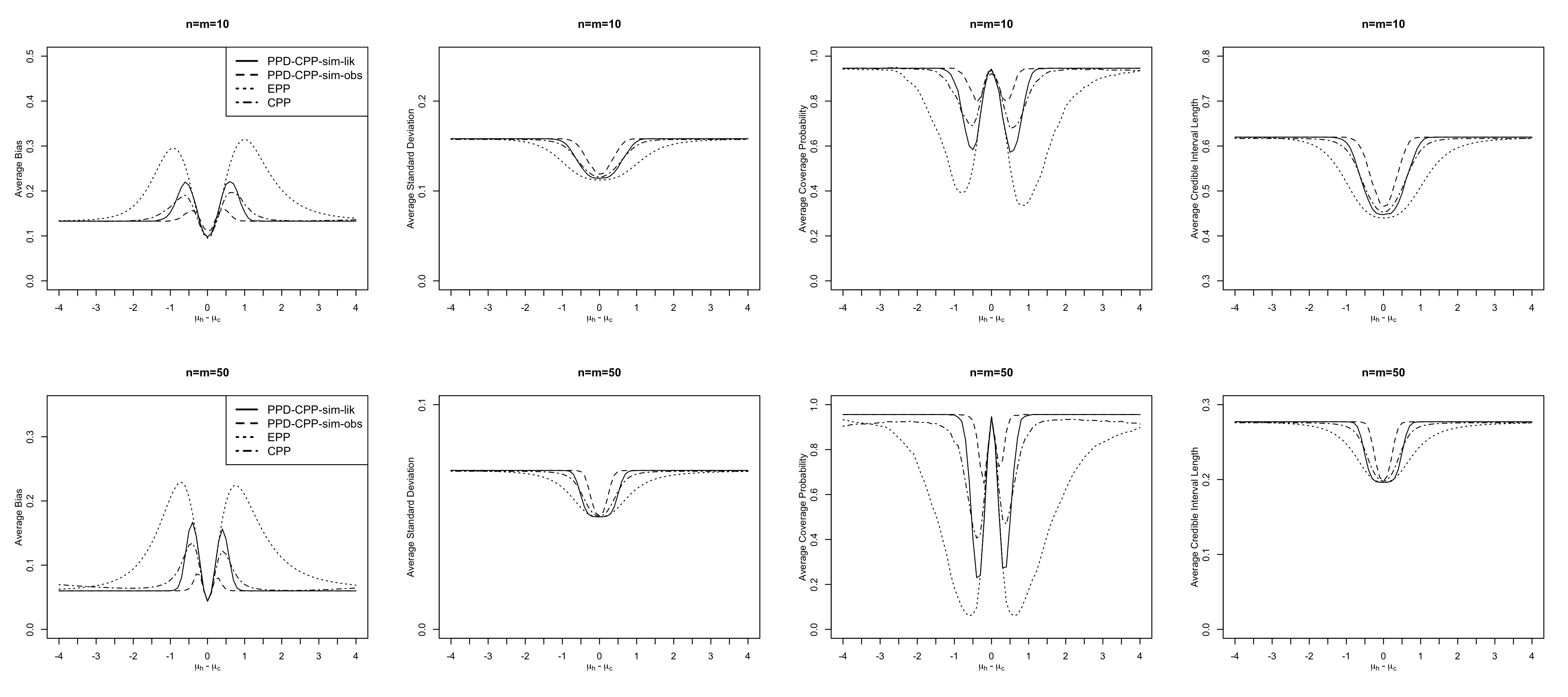}
            }
            \caption{Model performance for normal endpoints (no covariates) with known $\sigma^2_h$ and $\sigma^2_c$, using different power parameter determination methods. The posterior summary for each point on the curve is computed using $500$ simulation replicates.}
            \label{fig:3}
\end{figure}

    We first illustrate how power parameter $\alpha$ varies with mean difference $\mu_h-\mu_c$ when the variance $\sigma^2_h$ and $\sigma^2_c$ are known scalars. As shown in Figure~\ref{fig:2}, all methods yield a value of $\alpha$ over $0.8$ when data are almost congruent (i.e. ${\lvert \mu_h-\mu_c \rvert}/{0.5} \leq 0.2$) but a closed-to-$0$ $\alpha$ when data are strongly incongruent (i.e. ${\lvert \mu_h-\mu_c \rvert}/{0.5} \geq 6$). This tendency is amplified as sample size $n,m$ get larger, which validates Theorem 2 numerically. As ${\lvert \mu_h-\mu_c \rvert}$ increases, the proposed PPD-CPP discards incongruent information more rapidly than both CPP and EPP. Such sensitivity to incongruence makes PPD-CPP more risk-averse in historical borrowing. PPD-CPP-sim-obs is the most conservative method compared to others. This is due to its choice of test statistic in (1), which uses the observation itself  (i.e. $T(x)=x$), and is therefore statistically less informative than comparing likelihoods based on a correctly specified density function. It is also noteworthy that the proposed PPD-CPP exhibits higher probability of either completely borrowing or discarding historical information, which enhances the efficiency of analysis when data are compatible and reduces the risk of biasing inferences when data present growing conflicts.
    Figure~\ref{fig:3} summarizes the model performance for normal endpoints with known variance. The inverse bell-shaped curves for average posterior standard deviation and credible interval length illustrates that incorporating historical data reduces inferential uncertainty. Since PPD-CPP is sensitive to incongruence, they produce lower bias but higher coverage probability than EPP and CPP when ${\lvert \mu_h-\mu_c \rvert}/{0.5} \geq 2$. The conservative borrowing flavor of PPD-CPP-sim-obs yields the lowest bias and highest coverage probability when $n=m=50$. We notice that EPP and PPD-CPP-sim-lik display poor coverage probabilities around ${\lvert \mu_h-\mu_c \rvert}/{0.5} = 1$, where the corresponding average power parameter lies in the range of $(0.3,0.7)$. This is the region where historical and current data appear neither clearly congruent nor clearly incongruent and thus require dynamic borrowing. Figure~\ref{fig:3} shows that PPD-CPP-sim-lik outperforms EPP in both coverage probability and bias in this transitional zone. In practice, we can set $k_1=0$ to derive a more conservative borrowing behavior. The comparison between ``sim"-based and ``thm"-based PPD-CPP, along with a table presenting the numerical results for the noBorrow and complete pooling approaches, is provided in the Appendix~\ref{app:a6}. 

    Table~\ref{tab:1} displays the model performance for normal linear regression. The proposed PPD-CPP achieves comparable and in many cases superior bias and coverage probability relative to noBorrow and pooling method across different regression setups by effectively capturing the correct level of congruence. In most scenarios, ``obs"-based PPD-CPP yields a smaller power parameter estimates than ``lik"-based method. This results are expected, as in the presence of covariates, comparing raw observations in (1) may not be sufficiently informative in assessing the congruence and therefore may inflate bias (i.e. the case $\beta^h_0=49.5,\beta^h_0=50$). In this regard, we recommend ``lik"-based PPD-CPP in practice when covariates are available. For ``pw"-based PPD-CPP, which assigns pointwise power parameter for each historical observation, it introduces greater risks of inflating the estimation bias compared to the method who assigns a global power parameter. One interesting exception is that the proposed PPD-CPP-pw-lik could yield even smaller bias than noBorrow when $\beta^h_1=0, \beta^c_1=8$, where the historical binary covariate is inactive. This is because PPD-CPP-pw-lik successfully borrowed information from the historical reference group (i.e. $x_{1i}^h=0$), where congruence remains valid and the power parameter could reach $1$ for partial historical data. We conduct further simulation to study this phenomenon and the results are presented in the Appendix~\ref{app:a6}. The results of the rest simulation setups can also be found in the Appendix~\ref{app:a6}.

\begin{table}
    \centering
    \tiny
    \caption{Model performance for normal linear regression. The column Power denotes the average power parameter. Since ``pw"-based method assign each historical observation a unique power parameter, the 3-elements-vector, for example (0.88, 0.99, 1.00) denotes the average minimum, median, and maximum of $m$ power parameters from $500$ replicates. ``-" denotes the results are identical to noBorrow in 2 decimal digits.}
    
    \begin{tabular}{lcccccccccccccccc}
    
        \toprule
        \multirow{3}{*}{Setup} & \multirow{2}{*}{Method} & \multirow{2}{*}{Power} & 
        \multicolumn{3}{c}{Avg Bias} & \multicolumn{3}{c}{Avg Std Dev} & 
        \multicolumn{3}{c}{Coverage Probability} & \multicolumn{3}{c}{Avg Interval Length} \\
        \cmidrule(lr){4-6} \cmidrule(lr){7-9} \cmidrule(lr){10-12} \cmidrule(lr){13-15} 
        & & & $\beta_0$ & $\beta_1$ & $\beta_2$ & $\beta_0$ & $\beta_1$ & $\beta_2$ 
        & $\beta_0$ & $\beta_1$ & $\beta_2$ & $\beta_0$ & $\beta_1$ & $\beta_2$  \\
        \midrule
        Congruent   & PPD-CPP-thm-lik  & 0.98 & 0.25 & 0.08 & 0.004 & 0.33 & 0.10 & 0.01 & 0.95 & 0.96 & 0.94  & 1.28 & 0.40 & 0.02  \\
                    & PPD-CPP-thm-obs  & 0.98 & 0.25 & 0.08 & 0.004 & 0.33 & 0.10 & 0.01 & 0.94 & 0.96 & 0.93  & 1.28 & 0.40 & 0.02  \\
                    & PPD-CPP-sim-lik      & 0.97 & 0.25 & 0.08 & 0.01 & 0.33 & 0.10 & 0.01 & 0.95 & 0.96 & 0.94  & 1.28 & 0.40 & 0.02  \\
                    & PPD-CPP-sim-obs      & 0.96 & 0.25 & 0.08 & 0.004 & 0.33 & 0.10 & 0.01 & 0.94 & 0.96 & 0.93  & 1.29 & 0.41 & 0.02  \\
                    & PPD-CPP-pw-lik    & (0.88, 0.99, 1.00)   & 0.26 & 0.08 & 0.01 & 0.33 & 0.10 & 0.01 & 0.95 & 0.95 & 0.94  & 1.75 & 0.55 & 0.03  \\
                    & PPD-CPP-pw-obs    & (0.50, 0.96, 1.00)  & 0.26 & 0.08 & 0.01 & 0.33 & 0.11 & 0.01 & 0.96 & 0.94 & 0.95  & 1.78 & 0.56 & 0.03  \\
                    & noBorrow     & 0.00 & 0.35 & 0.11 & 0.01 & 0.47 & 0.15 & 0.01 & 0.96 & 0.95 & 0.95  & 1.84 & 0.58 & 0.03  \\
                    & pool         & 1.00 & 0.25 & 0.08 & 0.004 & 0.32 & 0.10 & 0.01 & 0.94 & 0.96 & 0.94  & 1.27 & 0.40 & 0.02  \\
        \midrule
        $\beta^h_0=49.5$,\\ $\beta^c_0=50$ & PPD-CPP-thm-lik  & 0.47 & 0.33 & 0.10 & 0.01 & 0.42 & 0.13 & 0.01 & 0.95 & 0.95 & 0.94 & 1.66 & 0.52 & 0.03  \\
        
                & PPD-CPP-thm-obs & 0.05 & 0.34 & 0.11 & 0.01 & 0.46 & 0.15 & 0.01 & 0.96 & 0.95 & 0.95 & 1.82 & 0.57 & 0.03  \\
                
                & PPD-CPP-sim-lik & 0.51 & 0.34 & 0.10 & 0.01 & 0.42 & 0.13 & 0.01 & 0.94 & 0.95 & 0.94 & 1.64 & 0.52 & 0.03  \\
                
                & PPD-CPP-sim-obs & 0.02 & 0.35 & 0.11 & 0.01 & 0.46 & 0.15 & 0.01 & 0.96 & 0.95 & 0.95 & 1.83 & 0.58 & 0.03  \\
                
                & PPD-CPP-pw-lik & (0.12, 0.49, 0.87) & 0.47 & 0.15 & 0.01 & 0.40 & 0.13 & 0.01 & 0.83 & 0.84 & 0.85 & 2.16 & 0.67 & 0.04  \\
                
                & PPD-CPP-pw-obs & (0.01, 0.06, 0.46) & 0.33 & 0.10 & 0.01 & 0.45 & 0.14 & 0.01 & 0.96 & 0.97 & 0.95 & 2.45 & 0.77 & 0.04  \\
                
                & noBorrow & 0.00 & 0.35 & 0.11 & 0.01 & 0.47 & 0.15 & 0.01 & 0.96 & 0.95 & 0.95 & 1.84 & 0.58 & 0.03  \\
                & pool & 1.00 & 0.34 & 0.09 & 0.01 & 0.36 & 0.11 & 0.01 & 0.91 & 0.95 & 0.95 & 1.42 & 0.48 & 0.03  \\
        \midrule
        $\beta^h_0=40$,\\ $\beta^c_0=50$ & PPD-CPP-thm-lik  & 0.00 & - & - & - & - & - & - & - & - & - & - & - & -  \\
                & PPD-CPP-thm-obs & 0.00 & - & - & - & - & - & - & - & - & - & - & - & -  \\
                & PPD-CPP-sim-lik & 0.00 & - & - & - & - & - & - & - & - & - & - & - & -  \\
                & PPD-CPP-sim-obs & 0.00 & - & - & - & - & - & - & - & - & - & - & - & -  \\
                & PPD-CPP-pw-lik & (0, 0, 0)  & 0.35 & 0.11 & 0.01 & 0.47 & 0.15 & 0.01 & 0.97 & 0.96 & 0.94 & 2.57 & 0.79 & 0.04  \\
                & PPD-CPP-pw-obs & (0, 0, 0) & 0.35 & 0.11 & 0.01 & 0.47 & 0.15 & 0.01 & 0.97 & 0.96 & 0.94 & 2.57 & 0.79 & 0.04  \\
                & noBorrow & 0.00 & 0.35 & 0.11 & 0.01 & 0.47 & 0.15 & 0.01 & 0.96 & 0.95 & 0.95 & 1.84 & 0.58 & 0.03  \\
                & pool & 1.00 & 5.06 & 0.79 & 0.05 & 3.27 & 1.03 & 0.06 & 0.67 & 0.96 & 0.95 & 12.86 & 4.05 & 0.23  \\
        \midrule
        $\beta^h_1=0$,\\ $\beta^c_1=8$ & PPD-CPP-thm-lik  & 0.02 & 0.39 & 0.24 & 0.01 & 0.72 & 0.23 & 0.01 & 0.98 & 0.93 & 0.98 & 2.84 & 0.89 & 0.05  \\
                & PPD-CPP-thm-obs & 0.03 & 0.41 & 0.28 & 0.01 & 0.76 & 0.24 & 0.01 & 0.98 & 0.92 & 0.98 & 3.01 & 0.95 & 0.05  \\
                & PPD-CPP-sim-lik & 0.03 & 0.40 & 0.29 & 0.01 & 0.76 & 0.24 & 0.01 & 0.98 & 0.93 & 0.98 & 2.99 & 0.94 & 0.05  \\
                & PPD-CPP-sim-obs & 0.04 & 0.42 & 0.33 & 0.01 & 0.78 & 0.25 & 0.01 & 0.98 & 0.91 & 0.98 & 3.08 & 0.97 & 0.06  \\
                & PPD-CPP-pw-lik & (0.00, 0.45, 0.99) & { 0.30} & 0.10 & 0.01 & 0.37 & 0.13 & 0.01 & 0.96 & 0.96 & 0.93 & 2.00 & 0.66 & 0.04  \\
                & PPD-CPP-pw-obs & (0.00, 0.39, 0.99) & 1.57 & 2.48 & 0.03 & 2.00 & 0.63 & 0.04 & 0.97 & 0.02 & 0.95 & 10.78 & 3.39 & 0.19  \\
                & noBorrow & 0.00 & 0.35 & 0.11 & 0.01 & 0.47 & 0.15 & 0.01 & 0.96 & 0.95 & 0.95 & 1.84 & 0.58 & 0.03  \\
                & pool & 1.00 & 1.43 & 4.04 & 0.03 & 1.87 & 0.59 & 0.03 & 0.96 & 0.00 & 0.96 & 7.34 & 2.31 & 0.13  \\
        \midrule
        $\beta^h_2=0$,\\ $\beta^c_2=0.5$ & PPD-CPP-thm-lik  & 0.00 & 0.35 & 0.11 & 0.01 & 0.49 & 0.15 & 0.01 & 0.96 & 0.96 & 0.95 & 1.92 & 0.60 & 0.03  \\
                & PPD-CPP-thm-obs & 0.00 & 0.35 & 0.11 & 0.01 & 0.49 & 0.15 & 0.01 & 0.96 & 0.96 & 0.95 & 1.92 & 0.60 & 0.03  \\
                & PPD-CPP-sim-lik & 0.00 & 0.35 & 0.11 & 0.01 & 0.49 & 0.15 & 0.01 & 0.96 & 0.96 & 0.95 & 1.92 & 0.60 & 0.03  \\
                & PPD-CPP-sim-obs & 0.00 & 0.35 & 0.11 & 0.01 & 0.49 & 0.15 & 0.01 & 0.96 & 0.96 & 0.95 & 1.92 & 0.60 & 0.03  \\
                & PPD-CPP-pw-lik & (0, 0, 0) & 0.35 & 0.11 & 0.01 & 0.48 & 0.15 & 0.01 & 0.97 & 0.97 & 0.95 & 2.64 & 0.82 & 0.05  \\
                & PPD-CPP-pw-obs & (0, 0, 0) & 0.35 & 0.11 & 0.01 & 0.48 & 0.15 & 0.01 & 0.97 & 0.97 & 0.95 & 2.64 & 0.82 & 0.05  \\
                & noBorrow & 0.00 & 0.35 & 0.11 & 0.01 & 0.47 & 0.15 & 0.01 & 0.96 & 0.95 & 0.95 & 1.84 & 0.58 & 0.03  \\
                & pool & 1.00 & 7.21 & 2.19 & 0.26 & 9.09 & 2.86 & 0.16 & 0.96 & 0.96 & 0.65 & 35.70 & 11.25 & 0.63  \\
        \bottomrule
    \end{tabular}
    \label{tab:1}
\end{table}

\end{spacing}

\section{Real data analysis}\label{sec: real_data}  
\begin{spacing}{1}
    In this section, we apply PPD-CPP to two real data examples for illustrations. The first data is from Mother's Gift study \parencite{zaman2008}, a double-blinded randomized controlled trial conducted in Bangladesh from August 2004 to December 2005. During the study, pregnant women are randomized to receive inactivated influenza vaccine (coded as the baseline) or 23-valent pneumococcal polysaccharide vaccine. After the delivery, the infants are randomized to take pneumococcal conjugate vaccine (pcv, coded as the baseline) or Haemophilus influenzae type b conjugate vaccine (hib). One purpose of this study is to investigate the impact of infant's vaccine type on the total weight gain (response variable) over the study period. Since there are two study sites involved (denoted by U and G), we consider G as the baseline site (78 pairs of mother and infants) while borrowing the entire information from site U (77 pairs of mother and infants) rather than ``control" only, using the PPD-CPP-thm-lik. The predictors include the infant's gender, the growth of infant's age (in weeks) over the study period, mother's vaccine type, and infant's vaccine type. Their regression coefficients are $\beta_1$ to $\beta_4$ respectively. The prior of the Bayesian linear model is $\pi(\bfbeta,\sigma^2)\sim 1/\sigma^2$ where $\bfbeta=(\beta_0,\hdots,\beta_4)^\top$, $\beta_0$ denotes the intercept, and $\sigma^2$ is the variance term.

Table \ref{tab:2}. presents the posterior analysis of Mother's Gift study. After accounting for the covariates, both methods conclude that the infants would gain less weight over the study period when taking the hib vaccine. Notably, PPD-CPP results in a more precise inference on all parameters since congruent information from site U is borrowed. The power parameter $\alpha$ is $1.00$ estimated using Lemma \ref{lmma2}, which supports the full exchangeability between site U and G.

The second example is the \textit{Ceriodaphnia dubia} test, which studies the decline in the number of organisms' offspring with respect to $6$ toxicity doses (i.e. $0,0.25\%,0.5\%,1\%,2\%,4\%$). The dose level of $0\%$ is the control group and there are $10$ observations at each dose level. We analyze the reproduction data by fitting a dose-response curve through generalized linear regression. We assume the reproduction counts for $i^{th}$ dose group follow Poisson distribution with parameter $\mu_i$, where $\log (\mu_{i})=\beta_0+\beta_1{c_i}+\beta_2{c_i}^2$ and $c_i$ denotes the dose level. We choose the test conducted by MNEPAD lab in January 1992 as the current data and borrow information of the control group from the test in April 1991. Since the responses are counts and assumed as Poisson, a closed form of $p_{CM}$ is not available and therefore we use the computational alternatives in (\ref{pcm_compu}). We use PPD-CPP-sim-obs to maintain a conservative borrowing flavor considering the small sample size. 

Table \ref{tab:2} shows the posterior summary of the \textit{Ceriodaphnia dubia} test. The power parameter is $0.55$ reflecting moderate level of exchangeability between current and historical control. By partial borrowing, the parameter estimates have smaller variance and credible interval length.
\begin{table}[ht]
\centering
\begin{tabular}{lllcccc}
\toprule
Study & Method & Parameter & Mean & SD & 95\% CI & Interval Length \\
\midrule
\multirow{10}{*}{Mother's Gift}
& \multirow{5}{*}{\centering noBorrow} 
  & $\beta_0$ & -0.10 & 0.36 & (-1.31, 0.49) & 1.81 \\
& & $\beta_1$ & 0.62  & 0.16 & (0.04, 0.89) & 0.85 \\
& & $\beta_2$ & 0.17  & 0.01 & (0.11, 0.20) & 0.08 \\
& & $\beta_3$ & 0.20  & 0.17 & (-0.44, 0.49) & 0.93 \\
& & $\beta_4$ & -0.36 & 0.17 & (-0.93, -0.10) & 0.83 \\
\addlinespace
& \multirow{5}{*}{\centering PPD-CPP-thm-lik}
  & $\beta_0$ & 0.02  & 0.28 & (-0.90, 0.50) & 1.40 \\
& & $\beta_1$ & 0.47  & 0.13 & (0.05, 0.68) & 0.63 \\
& & $\beta_2$ & 0.17  & 0.01 & (0.12, 0.19) & 0.06 \\
& & $\beta_3$ & 0.02  & 0.13 & (-0.48, 0.23) & 0.71 \\
& & $\beta_4$ & -0.38 & 0.12 & (-0.81, -0.18) & 0.63 \\
\midrule
\multirow{6}{*}{\textit{C. dubia} Test}
& \multirow{3}{*}{\centering noBorrow}
  & $\beta_0$ & 3.32  & 0.05 & (3.15, 3.40) & 0.25 \\
& & $\beta_1$ & 0.55  & 0.16 & (-0.02, 0.82) & 0.84 \\
& & $\beta_2$ & 0.78  & 0.10 & (-1.10, -0.62) & 0.48 \\
\addlinespace
& \multirow{3}{*}{\centering PPD-CPP-sim-obs} 
  & $\beta_0$ & 3.30  & 0.04 & (3.15, 3.37) & 0.22 \\
& & $\beta_1$ & 0.58  & 0.14 & (-0.01, 0.83) & 0.83 \\
& & $\beta_2$ & -0.79 & 0.09 & (-1.10, -0.66) & 0.45 \\
\bottomrule
\end{tabular}
\caption{Combined posterior summaries for the Mother's Gift study and the \textit{Ceriodaphnia dubia} test under noBorrow and PPD-CPP methods. SD denotes the posterior standard deviation. CI denotes the 95\% credible interval.}
\label{tab:2}

\end{table}
\end{spacing}

\section{Discussion}\label{sec: disc}  
\begin{spacing}{1}
    
    In this paper, we have developed a new congruence measure $p_{CM}$ based on marginal posterior predictive $p$-value for normal endpoints. $p_{CM}$ is highly interpretable w.r.t measuring how well historical data can replicate the current data. Instead of purely relying on MCMC \parencite{kwiatkowski2024}, $p_{CM}$ has closed forms and theoretical guarantees of convergence under either congruence or incongruence. We prove that $p_{CM}$ is no longer uniformly distributed when data are congruent. Based on $p_{CM}$ and its theoretical properties, we develop PPD-CPP to dynamically borrow overall or individualized historical information. We also generalize the calibration process in CPP and make the borrowing pattern free of any data (but only depends on current sample size $n$). This progress is important in historical borrowing as it reduces the risks of biasing the inference due to doubly/overly use of data (i.e. we use historical data to identify the information borrowing pattern and the level of borrowing, and then fit models using the data again). 

\begin{figure}
            \centerline{
            \includegraphics[scale=0.15]{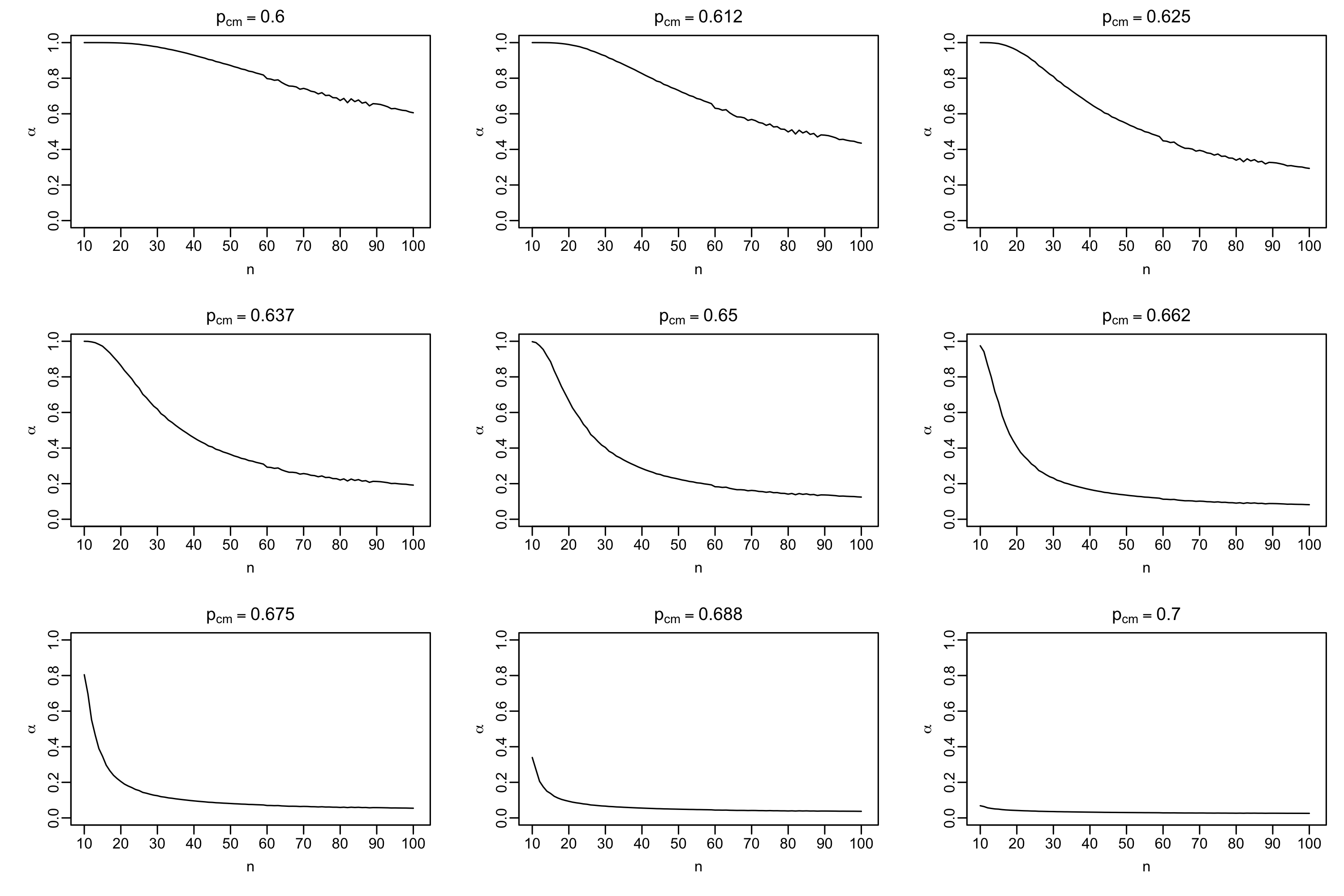}
            }
            \caption{Power parameter $\alpha$ vs the sample size $n$ of current control group when the true congruence level is fixed.}
            \label{fig:4}
\end{figure}

    PPD-CPP introduces several future research directions. One could be its generalization to Bernoulli, survival, or exponential family distributions. A closed form of $p_{CM}$ for these endpoints can produce a more accurate estimate of $\alpha$. The other one is in the realm of Bayesian adaptive designs. As shown in Figure~\ref{fig:4}, the power parameter is a monotone decreasing function of $n$ with an elbow point, a region where the rate of reduction in historical borrowing transitions from being rapid to gradual. In practice, this threshold may serve as a practical stopping criterion of sample size determinations, indicating when the current data are sufficiently informative for statistical inferences without further historical borrowing. We leave these topics for future investigation.

\end{spacing}

\printbibliography

\newpage
\renewcommand{\thesection}{A.\arabic{section}}

\section{Appendix}
\begin{spacing}{1}
    \subsection{$p_{CM}$ as a differential entropy representation}\label{app:a1}
    Here, we show that (\ref{pcm}) with posterior predictive likelihood (i.e. $p(x\mid \bfY^h)$) can be reformulated using differential entropy. Since $p(x\mid \bfY^h) \geq 0$,
        \begin{equation*}
        \begin{aligned}
             p_{CM} &=\Pr\left\{ T(y^{rep}_i) \geq  T(y^{c}_i) \mid \bfY^h \right\} \\
             & = \Pr\left\{ p(y^{rep}_i\mid \bfY^h) \geq  p(y^{c}_i\mid \bfY^h)  \right\} \\
             &= \Pr\left\{  -p(y^{rep}_i\mid \bfY^h) \log(p(y^{rep}_i\mid \bfY^h)) \leq -p(y^{rep}_i\mid \bfY^h)\log(p(y^{c}_i\mid \bfY^h))  \right\} \\
             &= \Pr\left\{ \int -p(y^{rep}_i\mid \bfY^h) \log(p(y^{rep}_i\mid \bfY^h)) dy^{rep}_i \leq -\log(p(y^{c}_i\mid \bfY^h))  \right\} \\
             &= \Pr\left\{  H(y^{rep}_i\mid \bfY^h)\leq -\log(p(y^{c}_i\mid \bfY^h))  \right\} \\
             &= \Pr\left\{  -H(y^{rep}_i\mid \bfY^h)\geq \log(p(y^{c}_i\mid \bfY^h))  \right\}
        \end{aligned}
    \end{equation*}
    $H(y^{rep}_i\mid \bfY^h)$ denotes the differential entropy of the posterior predictive samples $y^{rep}_i$.

    \subsection{Proof of Lemma $\ref{lmma}$ and Theorem $\ref{keythm}$}\label{app:a2}
    \subsubsection*{Normal endpoints with known variance}
    Let $y_i^h \sim N(\mu_h, \sigma_h^2), \, i = 1, \dots, m$ and $y_i^c \sim N(\mu_c, \sigma_c^2), \, i = 1, \dots, n$, where $\sigma_h^2$ and $\sigma_c^2$ are known. With flat prior $\pi(\mu_h) \sim 1$, the posterior predictive distribution is 
    $$
    y_i^{\text{rep}} \mid Y^h \sim N\left(\bar{Y}^h, \frac{m+1}{m} \sigma_h^2 \right), \, i = 1, \hdots, n
    $$
    Let $p(y_i^{{rep}} \mid \bfY^h)$ denote the posterior predictive density function for $y_i^{{rep}} \mid \bfY^h$. Here, the test statistic is $T(x;\bfY^h)=p(x\mid \bfY^h)$. Then,
    \begin{equation*}
        \begin{aligned}
            p_{CM} &=\Pr\left\{ T(y^{rep}_i) \geq  T(y^{c}_i) \mid \bfY^h \right\}\\
            &= \Pr \left\{ p(y_i^{{rep}} \mid \bfY^h) \geq p(y_i^c \mid \bfY^h) \right\} \\
            &= \Pr \left\{ 
                \frac{1}{\sqrt{2\pi \frac{m+1}{m} \sigma_h^2}} \exp \left[ -\frac{(y_i^{{rep}} - \bar{y}^h)^2}{2 \frac{m+1}{m} \sigma_h^2} \right]
                \geq 
                \frac{1}{\sqrt{2\pi \frac{m+1}{m} \sigma_h^2}} \exp \left[ -\frac{(y_i^c - \bar{y}^h)^2}{2 \frac{m+1}{m} \sigma_h^2} \right]
                \right\} \\
            &=\Pr \left\{ - (y_i^{{rep}} - \bar{y}^h)^2 \geq - (y_i^c - \bar{y}^h)^2 \right\} \\
            &= \Pr \left\{ (y_i^c - y_i^{{rep}})(y_i^c + y_i^{{rep}}) \geq 2 \bar{y}^h (y_i^c - y_i^{{rep}}) \right\} \\
            &= \Pr \left\{ y_i^c + y_i^{{rep}} - 2 \bar{y}^h \geq 0 \,\, \text{and} \,\, y_i^c - y_i^{{rep}} \geq 0 \right\} + \Pr \left\{ y_i^c + y_i^{{rep}} - 2 \bar{y}^h \leq 0 \,\, \text{and} \,\, y_i^c - y_i^{{rep}} \leq 0 \right\}\\
            &= \Pr \left\{ \begin{pmatrix} U \\ V \end{pmatrix} \geq \begin{pmatrix} 0 \\ 0 \end{pmatrix} \right\} + \Pr \left\{ \begin{pmatrix} U \\ V \end{pmatrix} \leq \begin{pmatrix} 0 \\ 0 \end{pmatrix} \right\}
        \end{aligned}
    \end{equation*}
    where $U = y_i^c + y_i^{{rep}} - 2\bar{y}^h$ and $V = y_i^c - y_i^{{rep}}.$ Note that in this case $\bar{y}^h$ is given, $y_i^{{rep}} \perp y_i^c$, and $U \sim N\left( \mu_c - \bar{y}^h, \, \sigma_c^2 + \frac{m+1}{m} \sigma_h^2 \right)$, $V \sim N\left( \mu_c - \bar{y}^h, \, \sigma_c^2 + \frac{m+1}{m} \sigma_h^2 \right)$ where $\text{cov}(U,V) = \sigma_c^2 - \frac{m+1}{m} \sigma_h^2$. Therefore,
    $$
    \begin{pmatrix} U \\ V \end{pmatrix} \Big | \;\bar{y}^h
        \sim \text{MVN} \left(
        \begin{pmatrix} \mu_c - \bar{y}^h \\ \mu_c - \bar{y}^h \end{pmatrix},
        \begin{bmatrix}
        \sigma_c^2 + \frac{m+1}{m} \sigma_h^2 & \sigma_c^2 - \frac{m+1}{m} \sigma_h^2 \\
        \sigma_c^2 - \frac{m+1}{m} \sigma_h^2 & \sigma_c^2 + \frac{m+1}{m} \sigma_h^2
    \end{bmatrix}
        \right).
    $$
    Asymptotically as \( m \to \infty \), by the weak law of large numbers ($\bar{y}^h\rightarrow \mu_h$ in probability) and slutsky theorem,
    $$
    \begin{pmatrix} U \\ V \end{pmatrix}
        \sim \text{MVN} \left(
        \begin{pmatrix} \mu_c - \mu_h \\ \mu_c - \mu_h \end{pmatrix},
        \begin{bmatrix}
        \sigma_c^2 + \sigma_h^2 & \sigma_c^2 - \sigma_h^2 \\
        \sigma_c^2 - \sigma_h^2 & \sigma_c^2 + \sigma_h^2
        \end{bmatrix}
        \right).
    $$
    \begin{enumerate}
        \item When data are congruent ($\mu_c = \mu_h, \sigma_c^2 = \sigma_h^2 = \sigma^2$), we have 
        $$
        \begin{pmatrix} U \\ V \end{pmatrix} \sim \text{MVN} \left( \bm{0}, \, 2\sigma^2 I_2 \right)
        $$
        and hence $p_{CM} = \frac{1}{4} + \frac{1}{4} = \frac{1}{2}.$

        \item When data are incongruent, we consider 3 cases of incongruence:
            \begin{itemize}
                \item Assume $\lvert \mu_c - \mu_h \rvert \to \infty  $ and $\frac{\sigma_c^2 + \sigma_h^2}{|\mu_c - \mu_h|} \to 0$, then $p_{CM} = 1.$ If we break apart the incongruence from $\lvert \mu_c - \mu_h \rvert \to \infty  $, then we have $p_{CM}=1$ as $ \mu_c - \mu_h  \to \infty  $, and $p_{CM}=0$ as $ \mu_c - \mu_h  \to -\infty  $. 
                \item Assume $\log\left( \frac{\sigma_h^2}{\sigma_c^2} \right) \to \infty$ (i.e. $\sigma_h^2 \gg \sigma_c^2$). This case is trivial since historical data will be white noise, and there is no need to borrow information from historical data. Note in this case the correlation coefficient $\rho =\frac{\text{cov} (U,V)}{\sigma_U\sigma_V} =\frac{\sigma^2_h-\sigma^2_h}{\sigma^2_c+\sigma^2_h}=-1$ and therefore $p_{CM}=0$.
                \item Assume $\log\left( \frac{\sigma_c^2}{\sigma_h^2} \right) \to \infty$ (i.e. $\sigma_c^2 \gg \sigma_h^2$). In this case, the correlation coefficient $\rho =\frac{\text{cov} (U,V)}{\sigma_U\sigma_V} =\frac{\sigma^2_h-\sigma^2_h}{\sigma^2_c+\sigma^2_h}=1$ and we have $p_{CM}=1$.
            \end{itemize}
    \end{enumerate}

    \subsubsection*{Normal endpoints with unknown variance}
    Let \( y_i^h \sim N(\mu_h, \sigma_h^2), \, i = 1, \dots, m \) and \( y_i^c \sim N(\mu_c, \sigma_c^2), \, i = 1, \dots, n \), where \( \sigma_h^2 \) and \( \sigma_c^2 \) are unknown. Let \( \pi(\mu_h, \sigma_h^2) \sim \frac{1}{\sigma_h^2} \) be the joint prior. Then, the posterior of $\mu_h$ and $\sigma^2_h$ is 
    \begin{equation*}
        \begin{aligned}
            p(\mu_h, \sigma_h^2 \mid \bfY^h) &\propto \left( \frac{1}{\sqrt{2\pi\sigma_h^2}} \right)^m 
            \exp \left\{ -\frac{\sum_{i=1}^m (y_i^h - \mu_h)^2}{2\sigma_h^2}  \right\} (\sigma_h^2)^{-1} \\
            &\propto (\sigma_h^2)^{-\frac{m+2}{2}} 
            \exp \left\{ -\frac{\sum_{i=1}^m (y_i^h - \mu_h)^2 + m (\mu_h - \bar{y}^h)^2}{2\sigma_h^2} \right\} \\
            &= (\sigma_h^2)^{-\frac{m+2}{2}} 
                \exp \left\{ -\frac{(m-1)S_h^2}{2\sigma_h^2} \right\} 
                \exp \left\{ -\frac{m(\mu_h - \bar{y}^h)^2}{2\sigma_h^2} \right\}
        \end{aligned}
    \end{equation*}
    where $S_h^2=\frac{\sum_{i=1}^m (y_i^h - \mu_h)^2}{m-1}$. Thus,
    \begin{equation*}
        \begin{aligned}
            p(\mu_h \mid \bfY^h, \sigma_h^2) &\propto \exp \left\{ -\frac{(\mu_h - \bar{y}^h)^2}{2\frac{\sigma_h^2}{m}} \right\} \\
            &\propto \mathcal{N}\left( \bar{y}^h, \frac{\sigma_h^2}{m} \right).
        \end{aligned}
    \end{equation*}
    Also, 
    \begin{equation*}
        \begin{aligned}
            p(\sigma_h^2 \mid \bfY^h) &\propto \frac{p(\mu_h, \sigma_h^2 \mid \bfY^h)}{p(\mu_h \mid \bfY^h, \sigma_h^2)} \\
            &\propto (\sigma_h^2)^{-\frac{m+2}{2}} (\sigma_h^2)^{\frac{1}{2}} 
            \exp \left\{ -\frac{(m-1)S_h^2 / 2}{\sigma_h^2} \right\}\\
            &= (\sigma_h^2)^{-\left( \frac{m}{2} + 1 \right)} 
            \exp \left\{ -\frac{(m-1)S_h^2 / 2}{\sigma_h^2} \right\}\\
            &\propto IG\left( \frac{m-1}{2}, \frac{(m-1)S_h^2}{2} \right)
        \end{aligned}
    \end{equation*}
    Therefore,
    $$
     p(\mu_h, \sigma_h^2 \mid \bfY^h) = {NIG} \left( \bar{y}^h, m, \frac{m-1}{2}, \frac{(m-1)S_h^2}{2} \right).
    $$
    By the conjugacy \parencite{murphy2007}
    \begin{equation*}
        \begin{aligned}
        p(y_i^{{rep}} \mid \bfY^h) &= \iint p(y_i^{{rep}} \mid \mu_h, \sigma_h^2) p(\mu_h, \sigma_h^2 \mid \bfY^h) d\mu_h d\sigma_h^2\\
        &\propto t_{(m-1)} \left( \bar{y}^h, \frac{m+1}{m}S_h^2 \right).
        \end{aligned}
    \end{equation*}
    Thus,
    \begin{equation*}
        \begin{aligned}
            p_{CM} &= \Pr \left\{ p(y_i^{{rep}} \mid \bfY^h) > P(y_i^c \mid \bfY^h) \right\} \\
            &= \Pr \Bigg\{ \frac{\Gamma\left( \frac{m}{2} \right)}{\Gamma\left( \frac{m-1}{2} \right) \sqrt{\pi(m-1)\frac{m+1}{m}S_h^2}} \left( 1 + \frac{1}{m-1} \frac{(y_i^{rep} - \bar{y}^h)^2}{\frac{m+1}{m}S_h^2} \right)^{-\frac{m}{2}}  \\ &\quad\quad\quad  \geq \frac{\Gamma\left( \frac{m}{2} \right)}{\Gamma\left( \frac{m-1}{2} \right) \sqrt{\pi(m-1)\frac{m+1}{m}S_h^2}} \left( 1 + \frac{1}{m-1} \frac{(y_i^{c} - \bar{y}^h)^2}{\frac{m+1}{m}S_h^2} \right)^{-\frac{m}{2}} \Bigg\}   \\
            &=\Pr \left\{ - (y_i^{{rep}} - \bar{y}^h)^2 \geq - (y_i^c - \bar{y}^h)^2 \right\} \\
            &= \Pr \left\{ (y_i^c - y_i^{{rep}})(y_i^c + y_i^{{rep}}) \geq 2 \bar{y}^h (y_i^c - y_i^{{rep}}) \right\} \\
            &= \Pr \left\{ y_i^c + y_i^{{rep}} - 2 \bar{y}^h \geq 0 \,\, \text{and} \,\, y_i^c - y_i^{{rep}} \geq 0 \right\} + \Pr \left\{ y_i^c + y_i^{{rep}} - 2 \bar{y}^h \leq 0 \,\, \text{and} \,\, y_i^c - y_i^{{rep}} \leq 0 \right\}\\
            &= \Pr \left\{ \begin{pmatrix} U \\ V \end{pmatrix} \geq \begin{pmatrix} 0 \\ 0 \end{pmatrix} \right\} + \Pr \left\{ \begin{pmatrix} U \\ V \end{pmatrix} \leq \begin{pmatrix} 0 \\ 0 \end{pmatrix} \right\}
        \end{aligned}
    \end{equation*}
    where $U = y_i^c + y_i^{{rep}} - 2\bar{y}^h$ and $V = y_i^c - y_i^{{rep}}.$ Note that in this case $\bar{y}^h,S^2_h$ are given and $y_i^{{rep}} \perp y_i^c$. Thus, as $m\to \infty$, $y^{rep}_i\sim \mathcal{N}(\bar{y}^h,S^2_h)$. Since $y^{c}_i\sim \mathcal{N}(\mu_c,\sigma^2_c)$, $U \sim N\left( \mu_c - \bar{y}^h, \, \sigma_c^2 + S^2_h \right)$, $V \sim N\left( \mu_c - \bar{y}^h, \, \sigma_c^2 + S^2_h \right)$ where $\text{cov}(U,V) = \sigma_c^2 -  S^2_h$.Therefore,
    $$
    \begin{pmatrix} U \\ V \end{pmatrix}
        \sim \mathcal{MVN} \left(
        \begin{pmatrix} \mu_c - \bar{y}^h \\ \mu_c - \bar{y}^h \end{pmatrix},
        \begin{bmatrix}
        \sigma_c^2 + S^2_h  & \sigma_c^2 - S^2_h \\
        \sigma_c^2 - S^2_h  & \sigma_c^2 + S^2_h 
    \end{bmatrix}
        \right).
    $$
    Asymptotically as \( m \to \infty \), by the weak law of large numbers,
    $$
    \begin{pmatrix} U \\ V \end{pmatrix}
        \sim \text{MVN} \left(
        \begin{pmatrix} \mu_c - \mu_h \\ \mu_c - \mu_h \end{pmatrix},
        \begin{bmatrix}
        \sigma_c^2 + \sigma_h^2 & \sigma_c^2 - \sigma_h^2 \\
        \sigma_c^2 - \sigma_h^2 & \sigma_c^2 + \sigma_h^2
        \end{bmatrix}
        \right).
    $$
    The discussion for the congruence and incongruence scenario is the same as the case for the normally distributed data with known variance.

    \subsubsection*{Normal endpoints with covariates (linear regression)}
    Let $y_i^h \sim N\left( {\mathbf{x}_i^{h}}^\top \bm{\beta}_h, \sigma_h^2 \right),i=1,\hdots,m$ and $y_i^c \sim N\left( {\mathbf{x}_i^{c}}^\top \bm{\beta}_c, \sigma_c^2 \right), i=1,\hdots,n$ where $\sigma_h^2$ and $\sigma_c^2$ are unknown. Let $\pi(\bm{\beta}_h, \sigma_h^2) \sim (\sigma_h^2)^{-\frac{p+2}{2}}$ be the Jeffrey prior where $p$ is the number of regression coefficients. The posterior distribution of $\bfbeta_h,\sigma^2_h$ is
    $$
    p(\bm{\beta}_h, \sigma_h^2 \mid \bfY^h) \propto (2\pi)^{-\frac{m}{2}} (\sigma_h^2)^{-\frac{m}{2}} \exp \left\{ -\frac{1}{2\sigma_h^2} \left( \bfY^h - X^h \bm{\beta}_h \right)^\top \left( Y^h - X^h \bm{\beta}_h \right) \right\} (\sigma_h^2)^{-\frac{p+2}{2}}
    $$
    where $X_h=\left ( {\mathbf{x}_1^{h}},\hdots,{\mathbf{x}_m^{h}}\right )^\top$. Therefore,
    \begin{equation*}
        \begin{aligned}
              p(\bm{\beta}_h \mid \bfY^h, \sigma_h^2) &\propto \exp \left\{ -\frac{1}{2\sigma_h^2} \left( \bfY^h - X^h \bm{\beta}_h \right)^\top \left( \bfY^h - X^h \bm{\beta}_h \right) \right\}\\
            &= \exp \left\{ -\frac{1}{2\sigma_h^2} \left( {\bfY^{h}}^\top \bfY^h - 2 \bm{\beta}_h^\top {X^{h}}^\top \bfY^h + \bm{\beta}_h^\top {X^{h}}^\top X^h \bm{\beta}_h \right) \right\} \\
            &\propto \mathcal{MVN} \left( ({X^{h}}^\top X^h)^{-1} {X^{h}}^\top \bfY^h, \, \sigma_h^2 ({X^{h}}^\top X^h)^{-1} \right).
        \end{aligned}
    \end{equation*}
    Also,
    \begin{equation*}
        \begin{aligned}
            p(\sigma_h^2 \mid \bfY^h) &\propto \frac{\pi(\bm{\beta}_h, \sigma_h^2 \mid \bfY^h)}{\pi(\bm{\beta}_h \mid \bfY^h, \sigma_h^2)} \\
            &\propto (\sigma_h^2)^{-\frac{m+p+2}{2}} (\sigma_h^2)^{\frac{p}{2}} \exp \left\{ -\frac{1}{2\sigma_h^2} \left( {\bm{\beta}^{*}}^\top  \bm{\beta}^* + {\bfY^{h}}^\top \bfY^h \right) \right\} \\
            &=(\sigma_h^2)^{-\frac{m+2}{2}} \exp \left\{ -\frac{1}{2\sigma_h^2} {\bfY^{h}}^\top \left( I_p - P \right) \bfY^h \right\} \\
            &\propto {IG}\left( \frac{m}{2}, \frac{{SSE}}{2} \right)
        \end{aligned}
    \end{equation*}
    where $\bm{\beta}^* = ({X^{h}}^\top X^h)^{-1} {X^{h}}^\top \bfY^h$ and $P = X^h ({X^{h}}^\top X^h)^{-1} {X^{h}}^\top.$ $I_p$ is a $p\times p$ identity matrix and $SSE={\bfY^{h}}^\top (I_p - P) \bfY^h$. Therefore,
    $$
    \pi(\bm{\beta}_h, \sigma_h^2 \mid \bfY^h) = {NIG} \left( \bm{\beta}^*, ({X^{h}}^\top X^h)^{-1}, \frac{m}{2}, \frac{{SSE}}{2} \right).
    $$
    By the conjugacy, we have 
    \begin{equation*}
        \begin{aligned}
            p(y_i^{{rep}} \mid \bfY^h) &= \iint p(y_i^{{rep}} \mid \bm{\beta}_h, \sigma_h^2) \pi(\bm{\beta}_h, \sigma_h^2 \mid \bfY^h) d\bm{\beta}_h d\sigma_h^2  \\
        &= t_m \left( {\mathbf{x}_i^{{rep}}}^\top \bm{\beta}^*, \frac{{SSE}}{m} \left( 1 + {\mathbf{x}_i^{{rep}}}^\top ({X^{h}}^\top X^h)^{-1} \mathbf{x}_i^{{rep}} \right) \right).
        \end{aligned}
    \end{equation*}
    Similar to the case $y_i^h \sim N(\mu_h, \sigma_h^2)$ where $\sigma_h^2$ is unknown,
    $$
     p_{CM} = \Pr \left\{ \begin{pmatrix} U \\ V \end{pmatrix} \leq \begin{pmatrix} 0 \\ 0 \end{pmatrix} \right\} + \Pr \left\{ \begin{pmatrix} U \\ V \end{pmatrix} \leq \begin{pmatrix} 0 \\ 0 \end{pmatrix} \right\}
    $$  where $U = y_i^c + y_i^{{rep}} - 2{\mathbf{x}_i^{{rep}}}^\top \bm{\beta}^*$ and $V = y_i^c - y_i^{{rep}}.$ Again, $y_i^{{rep}} \perp y_i^c$. Thus, as $m\to \infty$,  $y_i^{{rep}} \sim N\left( {\mathbf{x}_i^{{rep}}}^\top \bm{\beta}^*, \frac{{SSE}}{m} \left( 1 + {\mathbf{x}_i^{{rep}}}^\top ({X^{h}}^\top X^h)^{-1} \mathbf{x}_i^{{rep}} \right) \right)$ and $y_i^c \sim N\left( {\mathbf{x}_i^{c}}^\top \bm{\beta}_c, \sigma_c^2 \right).$ Therefore,
    $$
    \begin{pmatrix}
        U \\ V
    \end{pmatrix} \sim \mathcal{MVN} \left(
        \begin{pmatrix}
        {\mathbf{x}_i^{c}}^\top \bm{\beta}_c - {\mathbf{x}_i^{rep}}^\top \bm{\beta}^* \\
        {\mathbf{x}_i^{c}}^\top \bm{\beta}_c - {\mathbf{x}_i^{{rep}}}^\top \bm{\beta}^*
        \end{pmatrix},
        \begin{bmatrix}
        \sigma_c^2 + \sigma_h^2 H & \sigma_c^2 - \sigma_h^2 H \\
        \sigma_c^2 - \sigma_h^2 H & \sigma_c^2 + \sigma_h^2 H
        \end{bmatrix}
        \right)
    $$
    where $H = 1 + {\mathbf{x}_i^{{rep}}}^\top ({X^{h}}^\top X^h)^{-1} \mathbf{x}_i^{{rep}}$. In practice, we have $\mathbf{x}_i^{{rep}} = \mathbf{x}_i^c, \, i = 1, \dots, n$ and therefore as $m\to \infty$, by the consistency of the ordinary least square estimator,
    $$
     \begin{pmatrix}
        U \\ V
    \end{pmatrix}\sim \mathcal{MVN} \left(
        \begin{pmatrix}
        {\mathbf{x}_i^{c}}^\top (\bm{\beta}_c - \bm{\beta}_h) \\
        {\mathbf{x}_i^{c}}^\top (\bm{\beta}_c - \bm{\beta}_h)
        \end{pmatrix},
        \begin{bmatrix}
        \sigma_c^2 + \sigma_h^2 H_i & \sigma_c^2 - \sigma_h^2 H_i \\
        \sigma_c^2 - \sigma_h^2 H_i & \sigma_c^2 + \sigma_h^2 H_i
        \end{bmatrix}
        \right)
    $$
    where $H_i = 1 + {\mathbf{x}_i^{c}}^\top ({X^{h}}^\top X^h)^{-1} \mathbf{x}_i^c$. The proof of the linear regression case requires 2 assumptions shown in the followings:
    \subparagraph*{Assumption 1.}
    As \( m \to \infty \), \(\exists \, \mathbf{x}_i^h \in \mathbb{R}^p, \text{ s.t. } \mathbf{x}_i^h = \mathbf{x}_i^c \).
    \subparagraph*{Assumption 2.}
    $ X_h $ is full rank and $m>p$.
\\
\\
    Since $X_h$ is full rank, $\text{tr} \left( X^h ({X^{h}}^\top X^h)^{-1} {X^{h}}^\top \right) = \sum_{i=1}^m h_{ii} = p$ where $h_{ii} = \left[ \mathbf{x}_i^h ({X^{h}}^\top X^h)^{-1} {\mathbf{x}_i^{h}}^\top \right]$ is the leverage statistic. As $m\to\infty$, since $h_{ii} \in [0,1]$, $h_{ii} \to 0 $ to guarantee $\frac{1}{m} \sum h_{ii}$ is finite. By $\text{ Assumption 1.}$, 
    $$
    {\mathbf{x}_i^{c}}^\top ({X^{h}}^\top X^h)^{-1} \mathbf{x}_i^c \to 0, \quad i = 1, \dots, n.
    $$
    Therefore, we have $H_i \to 1,  i = 1, \dots, n.$ As a result,
    $$
    \begin{pmatrix} U \\ V \end{pmatrix}
        \sim \mathcal{MVN} \left(
        \begin{pmatrix}
        {\mathbf{x}_i^{c}}^\top (\bm{\beta}_c - \bm{\beta}_h) \\
        {\mathbf{x}_i^{c}}^\top (\bm{\beta}_c - \bm{\beta}_h)
        \end{pmatrix},
        \begin{bmatrix}
        \sigma_c^2 + \sigma_h^2 & \sigma_c^2 - \sigma_h^2 \\
        \sigma_c^2 - \sigma_h^2 & \sigma_c^2 + \sigma_h^2
        \end{bmatrix}
        \right).
    $$
    \begin{enumerate}
        \item When data are congruent ($\bfbeta_c = \bfbeta_h, \sigma_c^2 = \sigma_h^2 = \sigma^2$), we have 
        $$
        \begin{pmatrix} U \\ V \end{pmatrix} \sim \mathcal{MVN} \left( \bm{0}, \, 2\sigma^2 I_2 \right)
        $$
        and hence $p_{CM} = \frac{1}{4} + \frac{1}{4} = \frac{1}{2}.$

        \item When data are incongruent, we consider 4 cases of incongruence:
            \begin{itemize}
                \item Assume $\| \bfbeta_c - \bfbeta_h \|_p \to \infty  $ and $\frac{\sigma_c^2 + \sigma_h^2}{\| \bfbeta_c - \bfbeta_h \|_p} \to 0$, then $p_{CM} = 1.$ Note $\|\cdot\|_p$ denotes the $p-$norm.
                \item Assume $\log\left( \frac{\sigma_h^2}{\sigma_c^2} \right) \to \infty$ (i.e. $\sigma_h^2 \gg \sigma_c^2$). This case is trivial since historical data will be white noise, and there is no need to borrow information from historical data. Note in this case the correlation coefficient $\rho=-1$ and therefore $p_{CM}=0$.
                \item Assume $\log\left( \frac{\sigma_c^2}{\sigma_h^2} \right) \to \infty$ (i.e. $\sigma_c^2 \gg \sigma_h^2$). In this case, the correlation coefficient $\rho = 1$ and we have $p_{CM}=1$.
                \item Assume there exists $\bfx_i^c$ such that $H_i\to \infty$. This is the case where the covariate shift (a type of incongruence) comes into play and the assumption 1 is violated. In this case, $p_{CM}=0$ because the correlation coefficient $\rho=-1$.
            \end{itemize}
    \end{enumerate}

  \subsection{Closed form $p_{CM}$ with $T(x)=x$ and its asymptotic properties with or without covariates}\label{app:a3}
    \begin{lemma}\label{lmma_obs} 
        Let $\{y^h_i\}_{i=1:m}\simiid N(\mu_h,\sigma^2_h)$ and $\{y^c_i\}_{i=1:n} \simiid N(\mu_c,\sigma^2_c)$. $\sigma^2_h$ and $\sigma^2_c$ are known. Let the test statistic be the data observation itself, with $\pi_0(\mu_h)\propto 1$,
        \begin{equation*}
            p_{CM}=Pr \left\{ U\ge0 \right\} 
        \end{equation*}
        where $U=y^{rep}_i-y^c_i\sim N(\bar{y}^h-\mu_c,\sigma^2_c+\frac{m+1}m{\sigma^2_h})$. When $\sigma^2_h$ and $\sigma^2_c$ are unknown, with $\pi_0(\mu_h,\sigma^2_h)\propto \frac{1}{\sigma^2_h}$, $U=y^{rep}_i-y^c_i\sim N(\mu_h-\mu_c,\sigma^2_c+{\sigma^2_h})$ asymptotically.

        \begin{proof}
            Let $y_i^h \sim N(\mu_h, \sigma_h^2), \, i = 1, \dots, m$ and $y_i^c \sim N(\mu_c, \sigma_c^2), \, i = 1, \dots, n$, where $\sigma_h^2$ and $\sigma_c^2$ are known. With flat prior $\pi(\mu_h) \sim 1$, the posterior predictive distribution is 
    $$
    y_i^{\text{rep}} \mid Y^h \sim N\left(\bar{y}^h, \frac{m+1}{m} \sigma_h^2 \right), \, i = 1, \hdots, n
    $$
    Here, the test statistic is $T(x)=x$. Then,
    \begin{equation*}
        \begin{aligned}
            p_{CM} &=\Pr\left\{ T(y^{rep}_i) \geq  T(y^{c}_i) \mid \bfY^h \right\}\\
            &= \Pr\left\{ y^{rep}_i \geq  y^{c}_i \right\}\\
            &= \Pr\left\{ y^{rep}_i \geq  y^{c}_i \right\}\\
            &= \Pr\left\{ U\geq 0 \right\}
        \end{aligned}
    \end{equation*}
    where $y^{rep}_i \bot y^c_i$ and $U =y_i^{{rep}}- y_i^c\sim N(\bar{y}^h-\mu_c,\sigma^2_c+\frac{m+1}m{\sigma^2_h})$.  As $m\to \infty$, $U =y_i^{{rep}}- y_i^c\sim N(\mu_h-\mu_c,\sigma^2_c+{\sigma^2_h})$.
    \begin{enumerate}
        \item When data are congruent ($\mu_c = \mu_h, \sigma_c^2 = \sigma_h^2 = \sigma^2$), we have 
        $$
        U \sim N\left( {0}, \, 2\sigma^2  \right)
        $$
        and hence $p_{CM} =  \frac{1}{2}.$

        \item When data are incongruent, we consider 3 cases of incongruence:
            \begin{itemize}
                \item Assume $\lvert \mu_c - \mu_h \rvert \to \infty  $ and $\frac{\sigma_c^2 + \sigma_h^2}{|\mu_c - \mu_h|} \to 0$, then $p_{CM} = 1.$ 
                \item Assume $\log\left( \frac{\sigma_h^2}{\sigma_c^2} \right) \to \infty$ (i.e. $\sigma_h^2 \gg \sigma_c^2$) or $\log\left( \frac{\sigma_c^2}{\sigma_h^2} \right) \to \infty$ (i.e. $\sigma_c^2 \gg \sigma_h^2$). In such cases, $p_{CM}=\frac{1}{2}$ in this case which means $p_{CM}$ is unable to catch the data incongruence that is stemmed from variance difference. This is one limitation of choosing $T(x)=x$.
            \end{itemize}
    \end{enumerate}
    When $\sigma^2_c$ and $\sigma^2_h$ are unknown, with $\pi(\mu_h, \sigma_h^2) \sim \frac{1}{\sigma_h^2} $ be the joint prior, the posterior predictive distribution is 
    \begin{equation*}
        \begin{aligned}
        p(y_i^{{rep}} \mid \bfY^h) &= \iint p(y_i^{{rep}} \mid \mu_h, \sigma_h^2) p(\mu_h, \sigma_h^2 \mid \bfY^h) d\mu_h d\sigma_h^2\\
        &\propto t_{(m-1)} \left( \bar{y}^h, \frac{m+1}{m}S_h^2 \right).
        \end{aligned}
    \end{equation*}
    As $m\to \infty$, the $t$ distribution will become normal distributions with $S^2_h\to \sigma^2_h$, and therefore we have $Pr \left\{ U\ge0 \right\} $ where $U=y^{rep}_i-y^c_i\sim N(\mu_h-\mu_c,\sigma^2_c+{\sigma^2_h})$. The analysis under congruence and incongruence is the same as the case with known $\sigma^2_c$ and $\sigma^2_h$.
        \end{proof}
    \end{lemma}

    \begin{lemma}\label{lmma_obs_covariate} 
        Let $y_i^h \sim N\left( {\mathbf{x}_i^{h}}^\top \bm{\beta}_h, \sigma_h^2 \right),i=1,\hdots,m$ and $y_i^c \sim N\left( {\mathbf{x}_i^{c}}^\top \bm{\beta}_c, \sigma_c^2 \right), i=1,\hdots,n$ where $\mathbf{x}_i^{h},\mathbf{x}_i^{c}$ are the $p\times 1$ covariate vectors, $\bm{\beta}_h,\bm{\beta}_c$ are the $p\times 1$ vectors of regression parameters, $\sigma_h^2$ and $\sigma_c^2$ are unknown. $p$ is the number of regression coefficients. Let the test statistic be the data observation itself, with $\pi(\bm{\beta}_h, \sigma_h^2) \sim (\sigma_h^2)^{-\frac{p+2}{2}}$, the asymptotic closed form of the pointwise $p_{CM}$, $p_{CM,i}$, is
        \begin{equation*}
            p_{CM,i}=Pr \left\{ U_i \geq 0 \right\}
        \end{equation*}
       with  $
     U_i\sim \mathcal{N} \left(
        {\mathbf{x}_i^{c}}^\top (\bm{\beta}_c - \bm{\beta}_h),
        \sigma_c^2 + \sigma_h^2 H_i \right )
    $
    where $H_i = 1 + {\mathbf{x}_i^{c}}^\top ({X^{h}}^\top X^h)^{-1} \mathbf{x}_i^c$ and $i=1,\hdots,n$. $X^h$ is the $m\times p$ design matrix for historical data.

        \begin{proof}
            Let $y_i^h \sim N\left( {\mathbf{x}_i^{h}}^\top \bm{\beta}_h, \sigma_h^2 \right),i=1,\hdots,m$ and $y_i^c \sim N\left( {\mathbf{x}_i^{c}}^\top \bm{\beta}_c, \sigma_c^2 \right), i=1,\hdots,n$ where $\sigma_h^2$ and $\sigma_c^2$ are unknown. Let $\pi(\bm{\beta}_h, \sigma_h^2) \sim (\sigma_h^2)^{-\frac{p+2}{2}}$ be the Jeffrey prior where $p$ is the number of regression coefficients. By the proof in A.2 (regression case), we have 
            \begin{equation*}
        \begin{aligned}
            p(y_i^{{rep}} \mid \bfY^h) &= \iint p(y_i^{{rep}} \mid \bm{\beta}_h, \sigma_h^2) \pi(\bm{\beta}_h, \sigma_h^2 \mid \bfY^h) d\bm{\beta}_h d\sigma_h^2  \\
        &= t_m \left( {\mathbf{x}_i^{{rep}}}^\top \bm{\beta}^*, \frac{{SSE}}{m} \left( 1 + {\mathbf{x}_i^{{rep}}}^\top ({X^{h}}^\top X^h)^{-1} \mathbf{x}_i^{{rep}} \right) \right).
        \end{aligned}
    \end{equation*}
    By the definition of $p_{CM}$ and follow the proof in A.2, we have 
    \begin{equation*}
            p_{CM,i}=Pr \left\{ U_i \geq 0 \right\}
    \end{equation*}
    where $
     U_i=y^{rep}_i-y^c_i\sim \mathcal{N} \left(
        {\mathbf{x}_i^{c}}^\top (\bm{\beta}_c - \bm{\beta}_h),
        \sigma_c^2 + \sigma_h^2 H_i \right )
    $ as $m\to \infty$, and $H_i = 1 + {\mathbf{x}_i^{c}}^\top ({X^{h}}^\top X^h)^{-1} \mathbf{x}_i^c$. Note here $y^{rep}_i \bot y^c_i$.
    \subparagraph*{Assumption 1.}
    As \( m \to \infty \), \(\exists \, \mathbf{x}_i^h \in \mathbb{R}^p, \text{ s.t. } \mathbf{x}_i^h = \mathbf{x}_i^c \).
    \subparagraph*{Assumption 2.}
    $ X_h $ is full rank and $m>p$.
\\
\\
    With two assumptions shown above, we have 
    $$
    {\mathbf{x}_i^{c}}^\top ({X^{h}}^\top X^h)^{-1} \mathbf{x}_i^c \to 0, \quad i = 1, \dots, n.
    $$
    Therefore, $H_i \to 1,  i = 1, \dots, n$ and $
     U_i=y^{rep}_i-y^c_i\sim \mathcal{N} \left(
        {\mathbf{x}_i^{c}}^\top (\bm{\beta}_c - \bm{\beta}_h),
        \sigma_c^2 + \sigma_h^2  \right ).
    $
    \begin{enumerate}
        \item When data are congruent  ($\bfbeta_c = \bfbeta_h, \sigma_c^2 = \sigma_h^2 = \sigma^2$), we have 
        $$
        U_i \sim N\left( {0}, \, 2\sigma^2  \right)
        $$
        and hence $p_{CM} =  \frac{1}{2}.$

        \item When data are incongruent, we consider 3 cases of incongruence:
            \begin{itemize}
                \item Assume $\| \bfbeta_c - \bfbeta_h \|_p \to \infty  $ and $\frac{\sigma_c^2 + \sigma_h^2}{\| \bfbeta_c - \bfbeta_h \|_p} \to 0$, then $p_{CM} = 1.$
                \item Assume $\log\left( \frac{\sigma_h^2}{\sigma_c^2} \right) \to \infty$ (i.e. $\sigma_h^2 \gg \sigma_c^2$) or $\log\left( \frac{\sigma_c^2}{\sigma_h^2} \right) \to \infty$ (i.e. $\sigma_c^2 \gg \sigma_h^2$). In such cases, $p_{CM}=\frac{1}{2}$ which means $p_{CM}$ is unable to catch the data incongruence that is stemmed from variance difference. This is one limitation of choosing $T(x)=x$.
                \item  Assume there exists $\bfx_i^c$ such that $H_i\to \infty$. This is the case where the covariate shift (a type of incongruence) comes into play and the assumption 1 is violated. In this case, $p_{CM}=\frac{1}{2}$ which means $p_{CM}$ with $T(x)=x$ is unable to catch the data incongruence that is due to covariate shifts.
            \end{itemize}
    \end{enumerate}
        \end{proof}
    \end{lemma}

    \begin{theorem}\label{keythm_obs} 
    Let $\{y^h_i\}_{i=1:m}\simiid N(\mu_h,\sigma^2_h)$ and $\{y^c_i\}_{i=1:n} \simiid N(\mu_c,\sigma^2_c)$ be independent. Let test statistic be $T(x)=x$. Assume $\pi_0(\mu_h)\propto 1$ when $\sigma^2_h$ is known and  $\pi_0(\mu_h,\sigma^2_h)\propto \frac{1}{\sigma^2_h}$ when $\sigma^2_h$ are unknown. For known variance case, $\bftheta_h=\mu_h$ and $\bftheta_c=\mu_c$; for unknown variance case, $\bftheta_h=(\mu_h,\sigma^2_h)^\top$ and $\bftheta_c=(\mu_c,\sigma^2_c)^\top$. When historical data and current data are congruent (i.e. $\bftheta_h=\bftheta_c$), regardless of known or unknown variance, 
        $$
        p_{CM} = \frac{1}{2}
        $$
    as $m\rightarrow \infty$. When data present growing incongruence ($\lvert\mu_c-\mu_h\rvert \rightarrow\infty$),
    $$
    p_{CM} = 1
    $$
    as $m\rightarrow \infty$.
        \begin{proof}
            See proof of Lemma 4.
        \end{proof}
    \end{theorem}
    \begin{remark}
        With $T(x)=x$, $p_{CM}$ is unable to detect the incongruence that is due to variance difference (without covariates).
    \end{remark}
    \begin{remark}
        For linear regression case where the covariates are present, $p_{CM}$ is unable to detect the incongruence that is due to variance difference or covariate shifts. The proof can be found in the proof of Lemma 4. In words, $p_{CM}$ with $T(x)=x$ can capture the incongruence from the mean difference or regression coefficients difference, while being less sensitive to variance difference or covariates shift.
    \end{remark}

    \subsection{Proof of Theorem $\ref{2ndthm}$}\label{app:a4} 
    The two-parameter sigmoid function is given by
    $$
    \alpha = \frac{1}{1 + \exp \left\{ a + b \log g(p_{CM}) \right\}}
    $$
    where $g(p_{CM}) = \left| p_{CM} - \frac{1}{2} \right|$. $a\in \mathcal{R}, b>0$ are given.
    \subparagraph*{When current and historical data (i.e. observations in the control arm) are congruent,} 
    By Theorem $\ref{keythm}$, as $m\to \infty$, $p_{CM}=1$ which leads to $\left| p_{CM} - \frac{1}{2} \right|=0$. Therefore, $\alpha \to 0.$
    \subparagraph*{When current and historical data are incongruent,} regardless of the case of incongruence (i.e. they are incongruent either by mean difference or variance difference or covariates shift), by Theorem $\ref{keythm}$, $p_{CM}=1 \text{ or } 0$ and $\left| p_{CM} - \frac{1}{2} \right| = \frac{1}{2}$ as $m\to\infty$. Let $\alpha^C$ be a number close to 1 and $\alpha^{IC}$ be a number close to 0. The closed form solution for $a,b$ (Jiang et al. 2023) is 
    $$
    \begin{cases}
        a = \log \left( \frac{1 - \alpha^C}{\alpha^C} \right) - \frac{\log \left( \frac{(1 - \alpha^C)\alpha^{IC}}{(1-\alpha^{IC})\alpha^C}  \right) \log g_1 (p_{CM})}{\log g_1 (p_{CM}) - \log g_2 (p_{CM})}, \\[10pt]
        b = \frac{\log \left( \frac{(1-\alpha^{C}) \alpha^{IC}}{(1 - \alpha^{IC}) \alpha^{C}} \right)}{\log g_1 (p_{CM}) - \log g_2 (p_{CM})}.
    \end{cases}
    $$
    where $g_1(p_{CM}) = \left| \mathbb{E}(\frac{W^C}{n}) + k_1 - \frac{1}{2} \right|$ and $ g_2(p_{CM}) = \left| \mathbb{E}(\frac{W^{IC}}{n}) - k_2 - \frac{1}{2} \right|$. $k_1 = \max \left\{ \left| L_1 - \frac{1}{2} \right|, \left| U_1 - \frac{1}{2} \right| \right\} / \tau$ and $k_2 = \max \left\{ \left| L_2 - \frac{1}{2} \right|, \left| U_2 - \frac{1}{2} \right| \right\} / \tau.$ We let $\tau=2$ in practice. $L_i,U_i$ represent the lower and upper bound of the confidence interval for the proportion parameter of $W_i \sim {Binom}(n,p_i),i=1,2$ where $p_1=\frac{1}{2},p_2=1$ ($p_2$ can also be 0 and results in the same conclusion). We consider two commonly used confidence interval for $p_i$:  asymptotic confidence interval and Clopper-Pearson exact confidence interval. Let $w_i$ be the realization of random variable $W_i$.
    %(Brown et al. 2001)
    \begin{itemize}
        \item (Asymptotic) 
        \begin{equation*}
        \begin{aligned}
         L_i &= \frac{w_i}{n} -\frac{z_{\frac{\alpha}{2}}}{\sqrt{n}}\sqrt{\frac{w_i}{n}(1-\frac{w_i}{n})}\\
            U_i &= \frac{w_i}{n} +\frac{z_{\frac{\alpha}{2}}}{\sqrt{n}}\sqrt{\frac{w_i}{n}(1-\frac{w_i}{n})},\;\; i=1,2.
        \end{aligned}
    \end{equation*}
    where $z_{\frac{\alpha}{2}}$ is the $\frac{\alpha}{2}$ quantile of the standard normal distribution. As $n\to \infty$, $\frac{z_{\frac{\alpha}{2}}}{\sqrt{n}}\sqrt{\frac{w_i}{n}(1-\frac{w_i}{n})}\to0$. Then, $k_1\to0$ and $k_2\to \frac{1}{2}$. Therefore, $g_1(p_{CM}) \to 0$ and $g_2(p_{CM})\to \frac{1}{4}$.
    As a result, as $n,m\to \infty$, the sigmoid function under incongruence is given by
    \begin{equation*}
        \begin{aligned}
            a + b \log \frac{1}{2} &= \log  \frac{1 - \alpha^{C}}{\alpha^{C}} + \left[ \log \left( \frac{(1 - \alpha^{C}) \alpha^{IC}}{(1 - \alpha^{IC}) \alpha^{C}} \right) \right] \left[ \frac{\log \frac{1}{2} - \log g_1(p_{CM})}{\log g_1(p_{CM}) - \log g_2(p_{CM})} \right]\\
            &=\log \left( \frac{1 - \alpha^{C}}{\alpha^{C}} \right) + \log \left( \frac{(1 - \alpha^{C}) \alpha^{IC}}{(1 - \alpha^{IC}) \alpha^{C}} \right) \cdot (-1) \\
            &=\log \left[ \frac{1 - \alpha^{IC}}{\alpha^{IC}} \right] \to \infty\quad \text{as  } \alpha^{IC} \to 0.
        \end{aligned}
    \end{equation*}
    Therefore, $\alpha \to 0$. Note that we consider $m\to \infty$ because Theorem 1. allows us to make the assumption such that $W^C \sim {Binom}(n,p_1=\frac{1}{2})$ and $W^{IC} \sim {Binom}(n,p_2=1)$.
        \item (Clopper-Pearson)

        \begin{equation*}
        \begin{aligned}
            L_i &=  \left(1+ \frac{n - w_i + 1}{w_i F_{\frac{\alpha}{2}; 2w_i, 2(n - w_i + 1)}} \right)^{-1}\\
            U_i &= \left(1+ \frac{n - w_i}{(w_i + 1) F_{\frac{\alpha}{2}; 2(w_i + 1), 2(n - w_i)}} \right)^{-1},\quad i=1,2.
        \end{aligned}
    \end{equation*}
    By the weak law of large number, $\frac{W^C}{n} \overset{P}{\to} \frac{1}{2},   \frac{W^{IC}}{n} \overset{P}{\to} 1$, and therefore by continuous mapping theorem, 
    $$
    \frac{n - W^C + 1}{W^C} \overset{P}{\to} 1, \quad \frac{n - W^C}{W^C + 1} \overset{P}{\to} 1;\quad
    \frac{n - W^{IC} + 1}{W^{IC}} \overset{P}{\to} 0, \quad \frac{n - W^{IC}}{W^{IC} + 1} \overset{P}{\to} 0.
    $$
    When $i=1$, $F_{\frac{\alpha}{2}; 2W^C, 2(n - wa_1 + 1)} \to 1 $ and $F_{\frac{\alpha}{2}; 2(W^C + 1), 2(n - W^C)} \to 1$. Thus by Slutsky's theorem, $L_1 \overset{P}{\to} \frac{1}{2},  U_1 \overset{P}{\to} \frac{1}{2}$ and $k_1 \overset{P}{\to} 0$. Therefore, $g_1(p_{CM}) \overset{P}{\to} 0$.\\
    When $i=2$, $L_2 \overset{P}{\to} 1,  U_2 \overset{P}{\to} 1$ and $k_1 \overset{P}{\to} \frac{1}{2}$. Therefore, $g_2(p_{CM}) \overset{P}{\to} \frac{1}{4}$. Then we can derive $\alpha \to 0$ as $\alpha^{IC} \to 0$ similar to the above proof using asymptotic confidence interval.\\
    \end{itemize}

    \subsection{Uniformity of $p_{CM}$ under the null (i.e. historical and current data are congruent) using (\ref{pcm_1})}\label{app:a5}
    \begin{figure}
            \centerline{
            \includegraphics[scale=0.15]{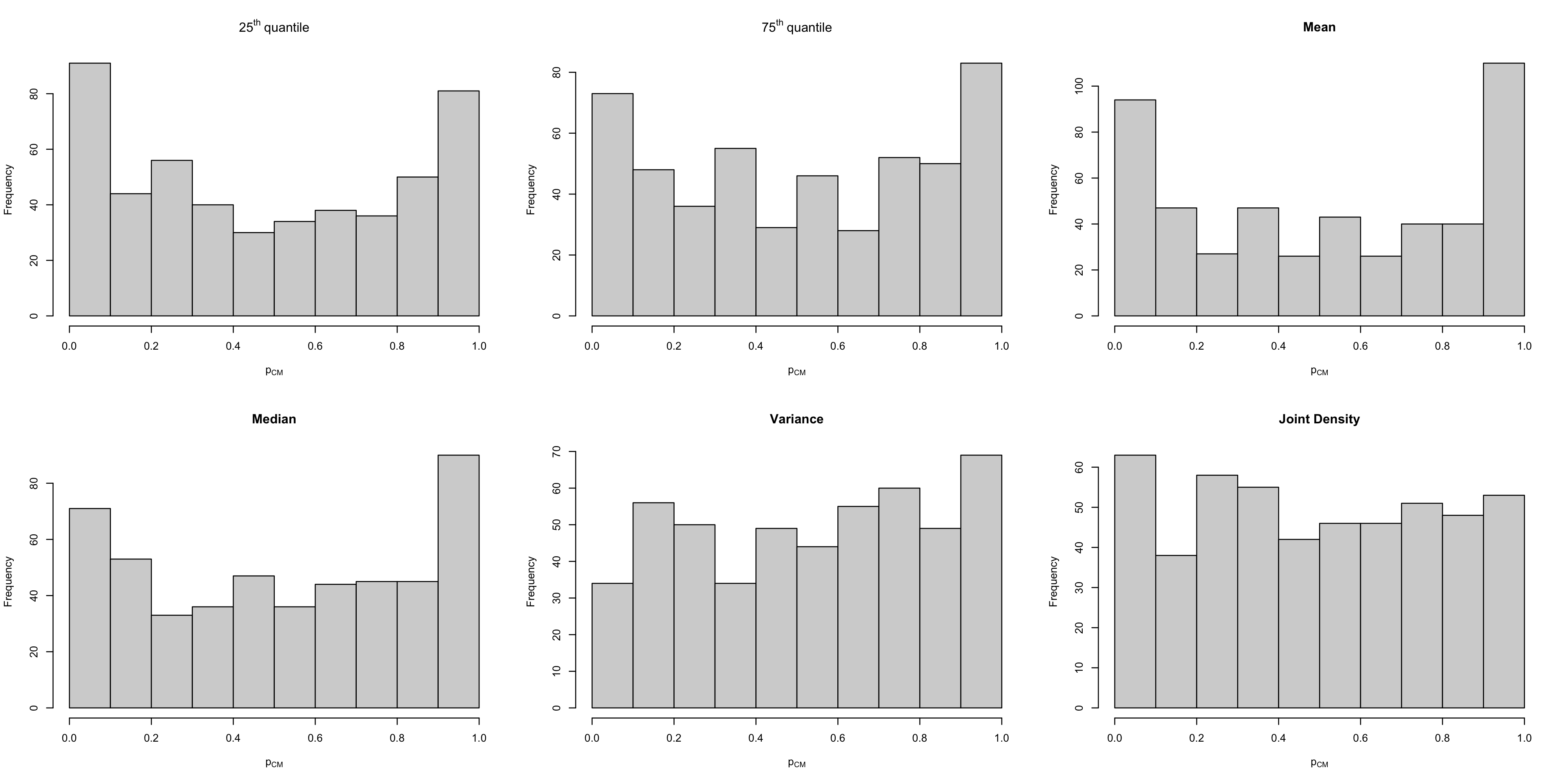}
            }
            \caption{The distribution of $p_{CM}$ defined in (\ref{pcm_1}) when historical and current data are congruent and both generated from $N(20,0.5^2)$ with sample size $50$. There are $500$ pairs of historical and current data are generated.}
            \label{fig:5}
\end{figure}

    \subsection*{Additional simulations}\label{app:a6}
    \subsubsection*{Mean difference (known variance)}
\begin{figure}
            \centerline{
            \includegraphics[scale=0.1]{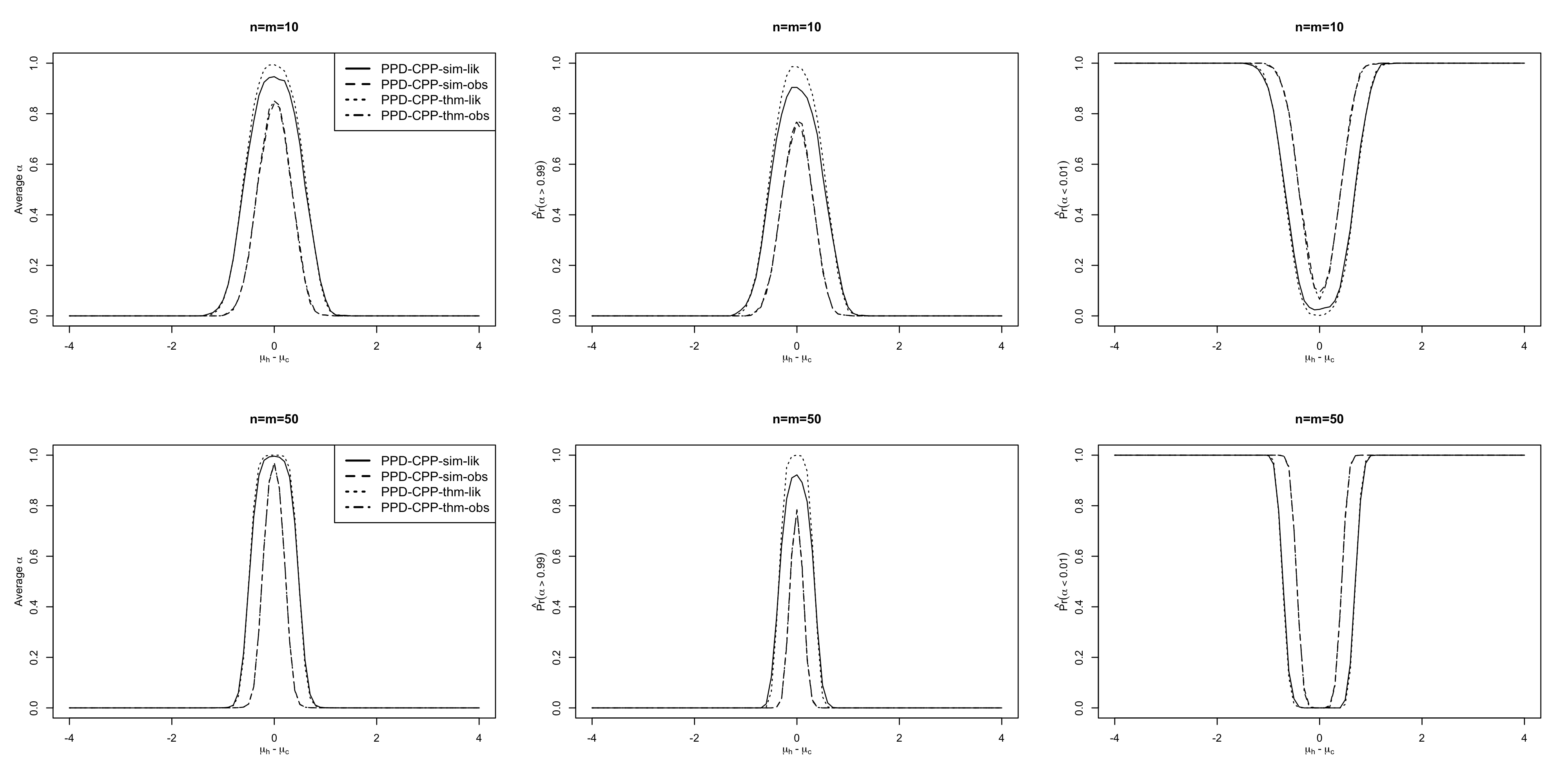}
            }
            \caption{Historical borrowing behavior comparison when $p_{CM}$ is derived either from Lemma \ref{lmma},\ref{lmma_obs} or computational method (\ref{pcm_compu}), with the change of mean difference for normal endpoints (no covariates), where $\sigma^2_h$ and $\sigma^2_c$ are assumed known.  The value on the curve is computed using $500$ power parameters.}
            \label{fig:6}
\end{figure}

\begin{figure}
            \centerline{
            \includegraphics[scale=0.1]{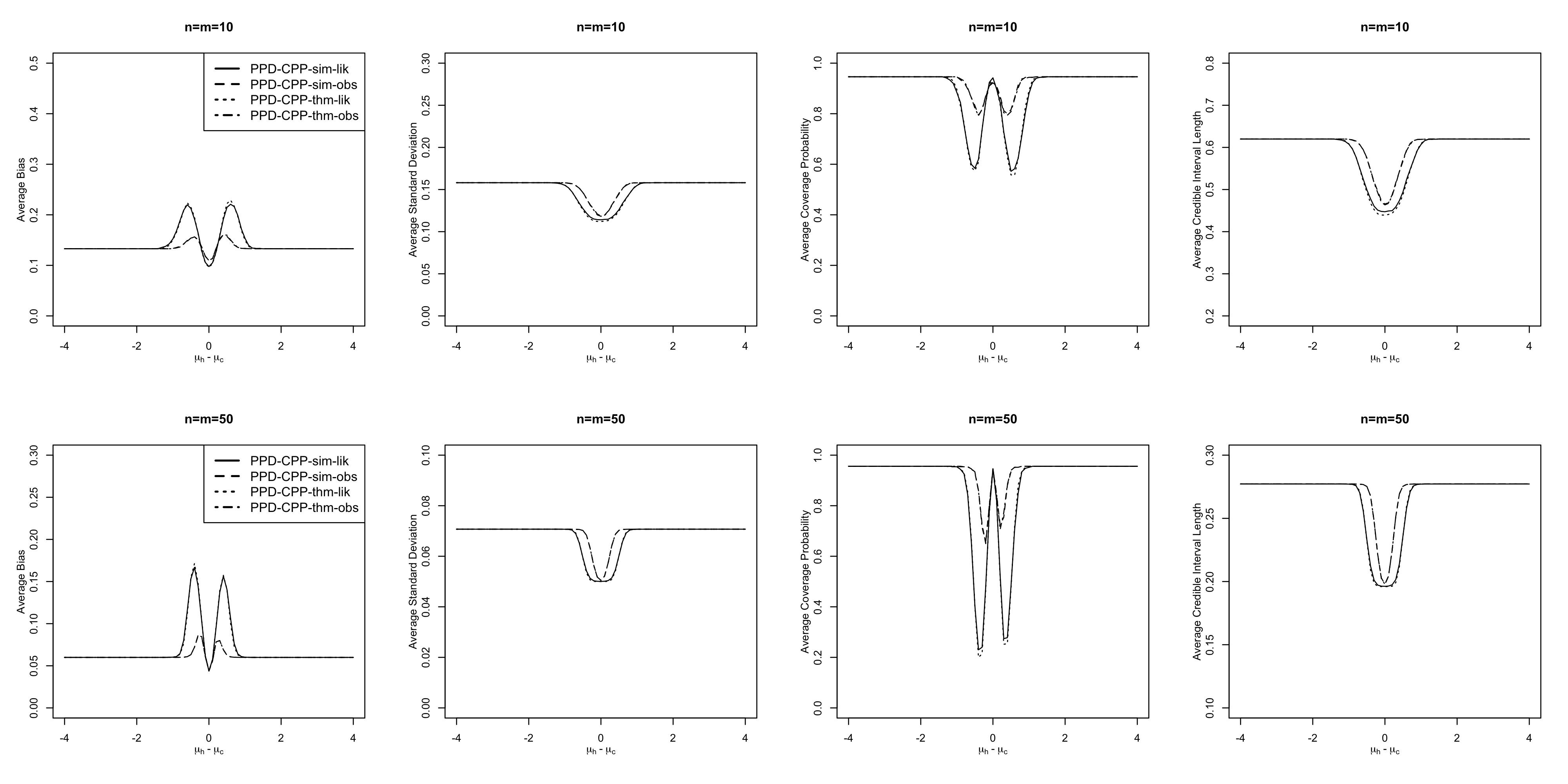}
            }
            \caption{Model performance comparison when $p_{CM}$ is derived either from Lemma \ref{lmma},\ref{lmma_obs} or computational method (\ref{pcm_compu}),for normal endpoints (no covariates) with known $\sigma^2_h$ and $\sigma^2_c$, using different power parameter determination methods. The posterior summary for each point on the curve is computed using $500$ simulation replicates. }
            \label{fig:7}
\end{figure}

    Figure~\ref{fig:6} and Figure~\ref{fig:7} compare the PPD-CPP when $p_{CM}$ is computed either by the theoretical closed-form expression (denoted as ``thm") or the computational approach (denoted by ``sim"). One notable observation is that when the test statistic is chosen as $T(x;\bfY^h)=p(x\mid\bfY^h)$ (denoted as ``lik"), PPD-CPP-sim-lik turns more conservative in historical borrowing than PPD-CPP-thm-lik for $n=m=10$. This is because extra variability is introduced by MCMC. As the sample size grows, this issue will be reduced.

    \subsubsection*{Mean difference (unknown variance)}
    This simulation adopts the same mean difference setup as discussed in the main text, but assumes the variance is unknown. Based on the figure, we observe a similar borrowing pattern to the case with known variance. Notably, there are slight differences between the borrowing behavior of PPD-CPP-sim-lik and PPD-CPP-thm-lik. However, these differences diminish as the sample size increases.
\begin{figure}
            \centerline{
            \includegraphics[scale=0.1]{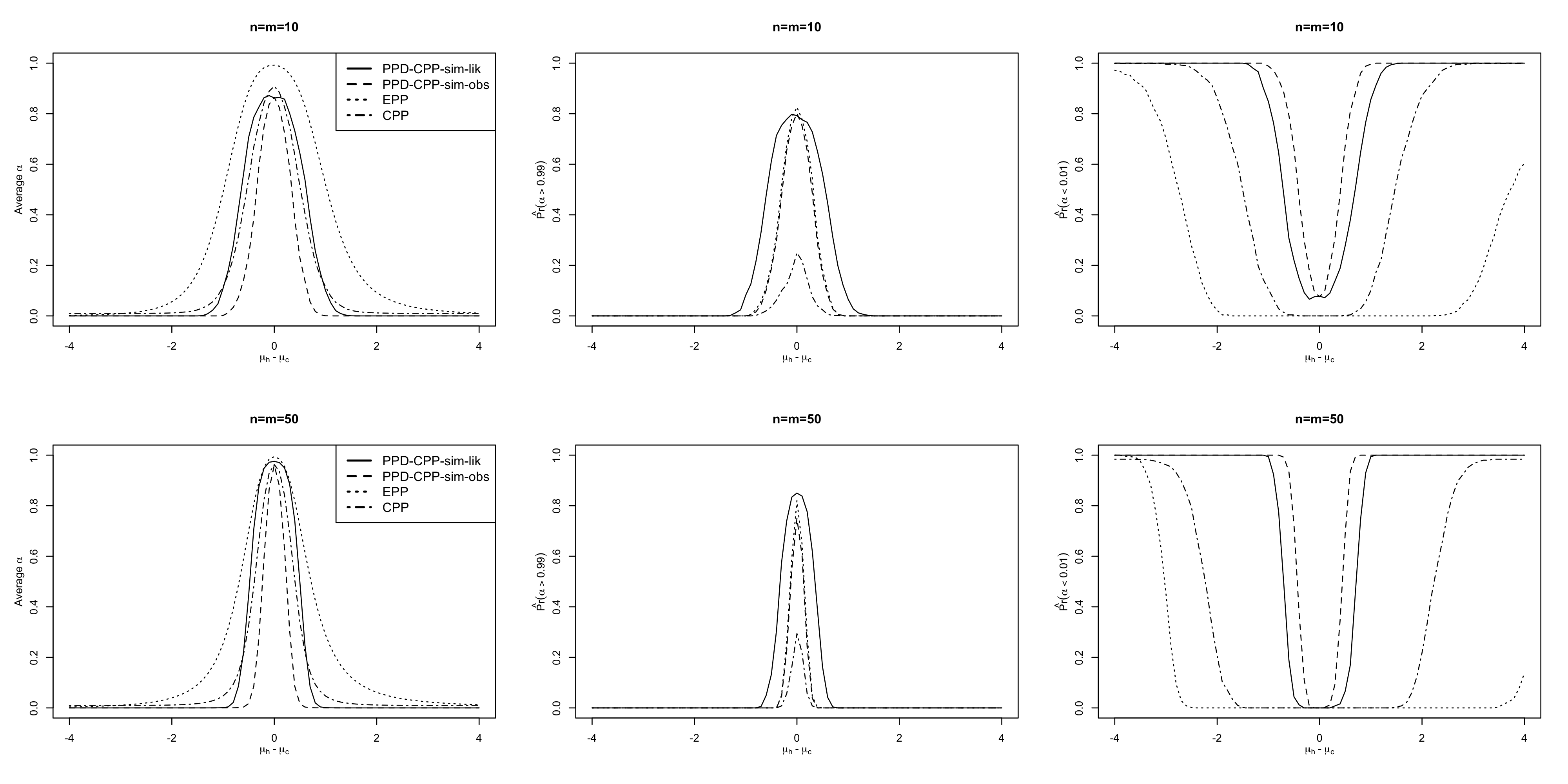}
            }
            \caption{Historical borrowing behavior with the change of mean difference for normal endpoints (no covariates), where $\sigma^2_h$ and $\sigma^2_c$ are assumed unknown. The value on the curve is computed using $500$ power parameters.}
            \label{fig:8}
\end{figure}
\begin{figure}
            \centerline{
            \includegraphics[scale=0.1]{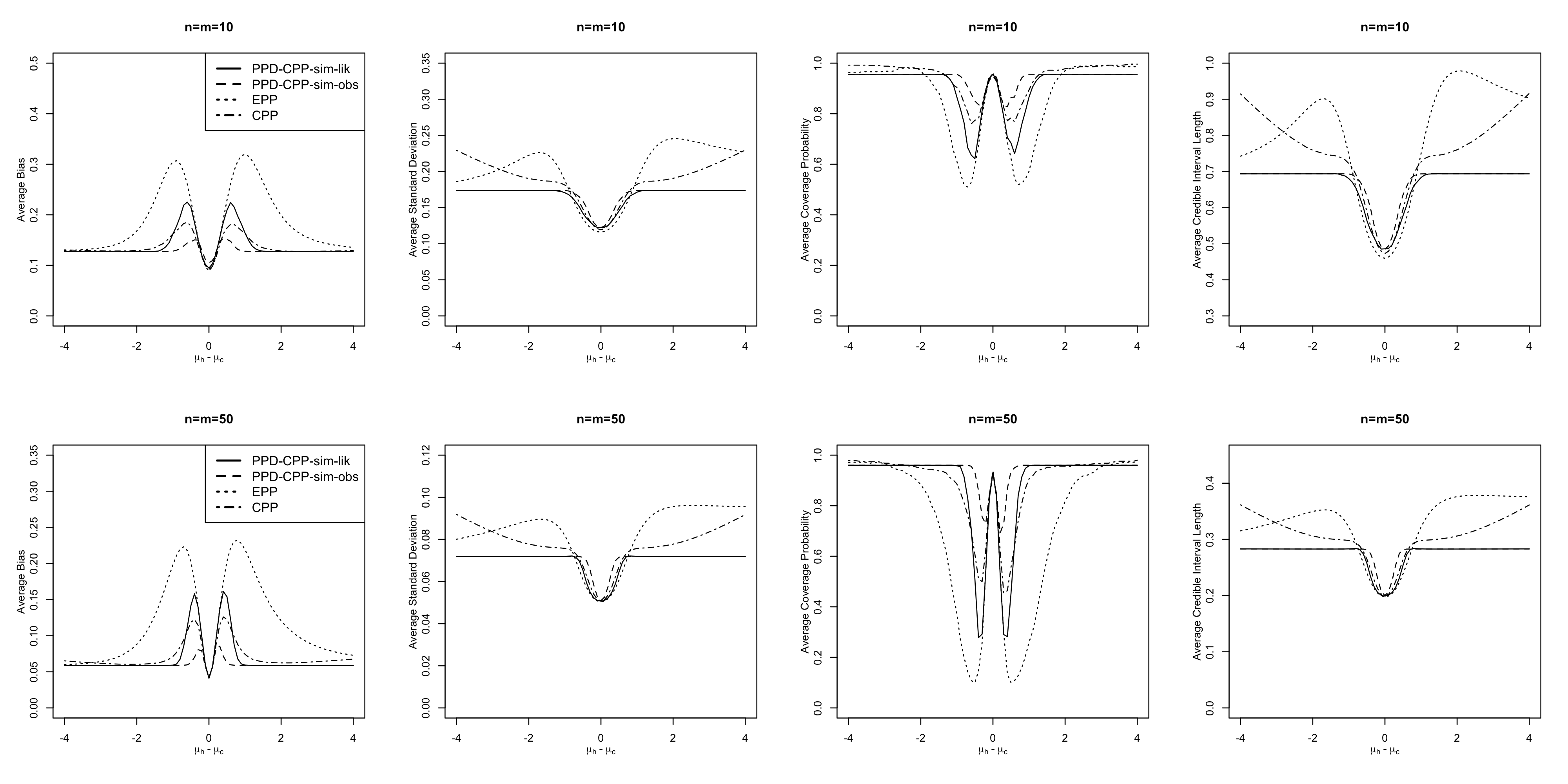}
            }
            \caption{Model performance for normal endpoints (no covariates) with unknown $\sigma^2_h$ and $\sigma^2_c$, using different power parameter determination methods. The posterior summary for each point on the curve is computed using $500$ simulation replicates.}
            \label{fig:9}
\end{figure}    
\begin{figure}
            \centerline{
            \includegraphics[scale=0.1]{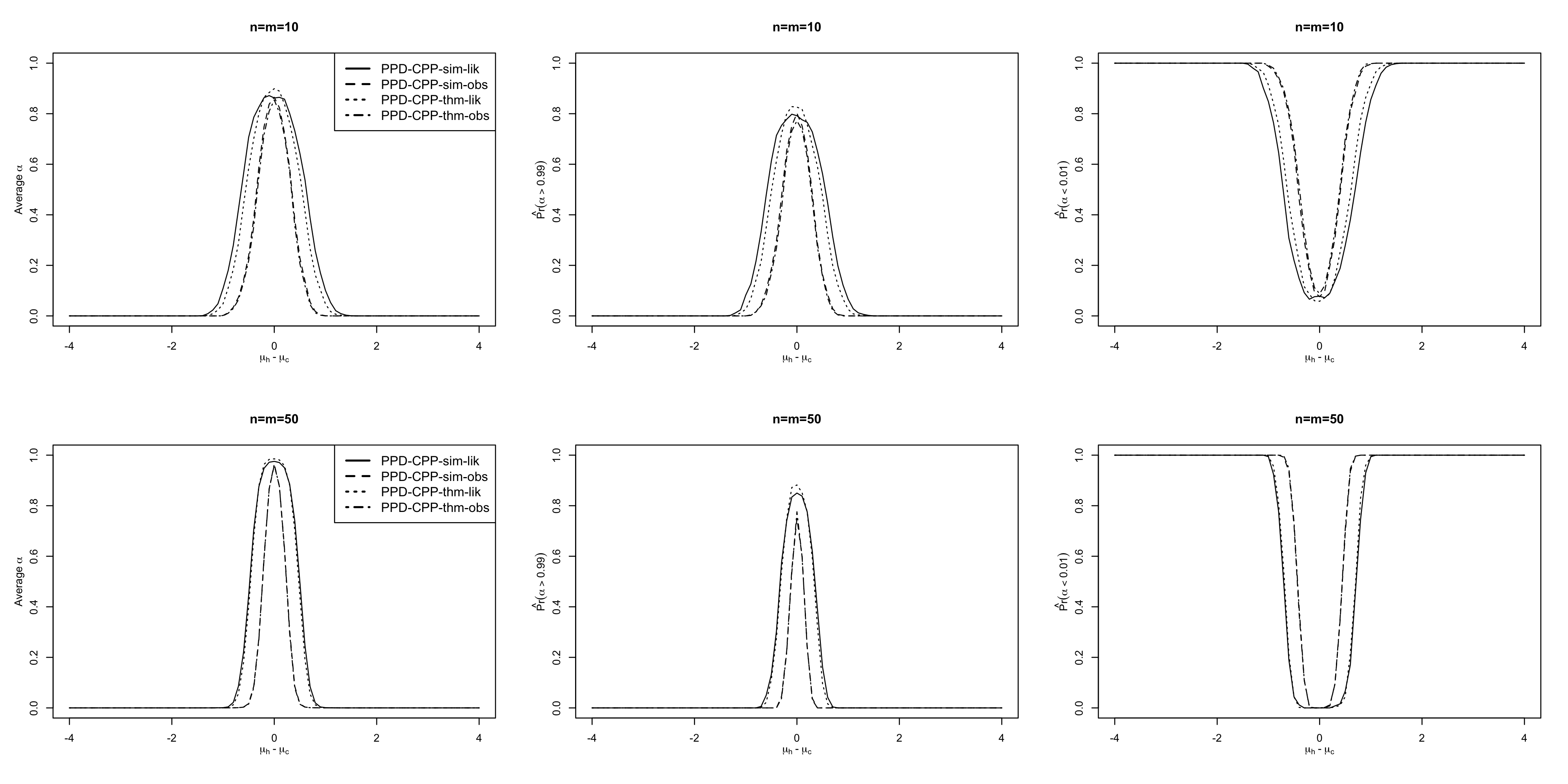}
            }
            \caption{Historical borrowing behavior comparison when $p_{CM}$ is derived either from Lemma \ref{lmma},\ref{lmma_obs} or computational method (\ref{pcm_compu}), with the change of mean difference for normal endpoints (no covariates), where $\sigma^2_h$ and $\sigma^2_c$ are assumed unknown.  The value on the curve is computed using $500$ power parameters.}
            \label{fig:10}
\end{figure}    
\begin{figure}
            \centerline{
            \includegraphics[scale=0.1]{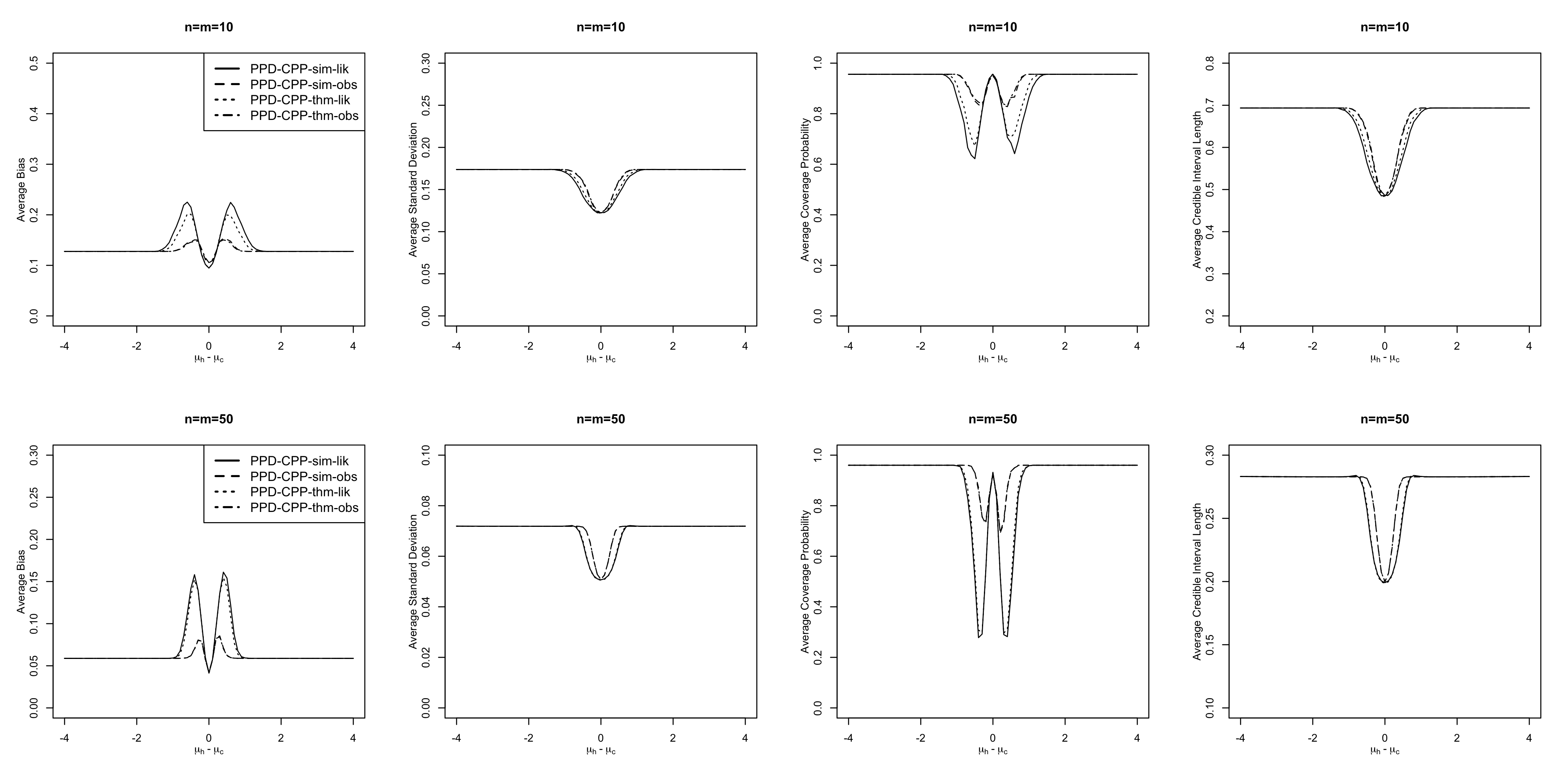}
            }
            \caption{Model performance comparison when $p_{CM}$ is derived either from Lemma \ref{lmma},\ref{lmma_obs} or computational method (\ref{pcm_compu}),for normal endpoints (no covariates) with unknown $\sigma^2_h$ and $\sigma^2_c$, using different power parameter determination methods. The posterior summary for each point on the curve is computed using $500$ simulation replicates. }
            \label{fig:11}
\end{figure}

    \subsubsection*{Mean difference when historical and current sample sizes are not equal (i.e. $n=10, m=40$)}

    We consider a simulation where the sample sizes for the historical and current data are \( m = 40 \) and \( n = 10 \) respectively. The incongruence stems from the mean difference, and the results are presented in the following figures. To prevent the analysis from being dominated by the historical information, a practical approach is to impose an upper bound on the power parameter, setting it to at most \( \frac{n}{m} \). In this case, we have \( \frac{n}{m} = \frac{10}{40} = 0.25 \) to ensure the effective contribution of the historical data does not exceed the sample size of the current data.

\begin{figure}
            \centerline{
            \includegraphics[scale=0.1]{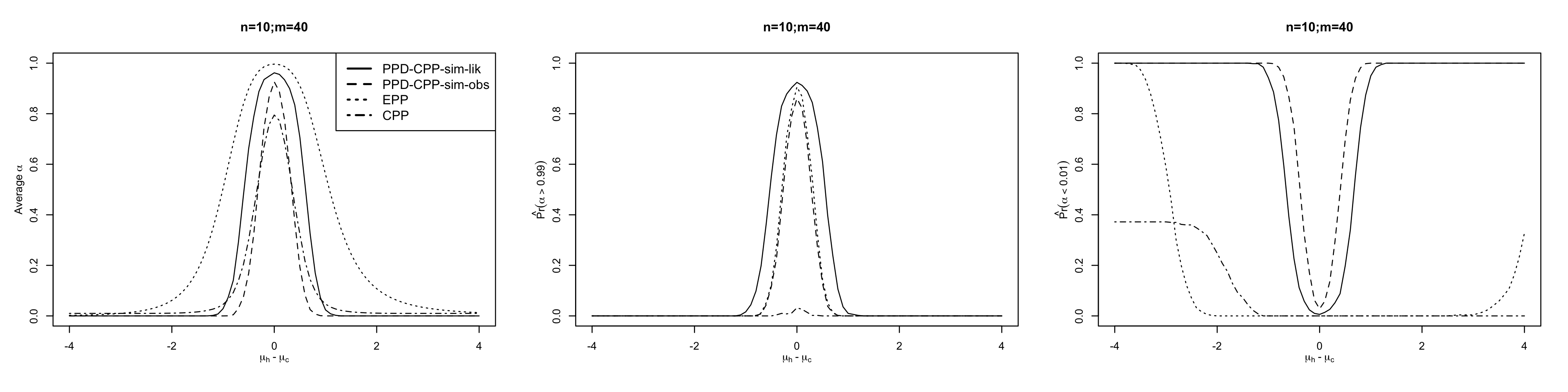}
            }
            \caption{Historical borrowing behavior with the change of mean difference for normal endpoints (no covariates), where $\sigma^2_h$ and $\sigma^2_c$ are assumed known. The value on the curve is computed using $500$ power parameters. Note in this case we have $n=10$ but $m=40$.}
            \label{fig:12}
\end{figure}
\begin{figure}
            \centerline{
            \includegraphics[scale=0.1]{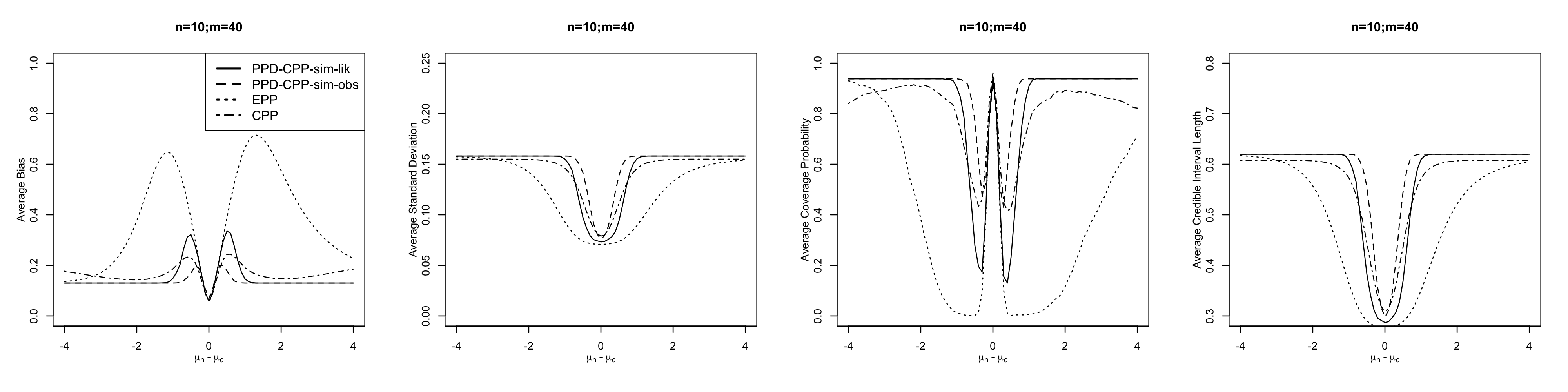}
            }
            \caption{Model performance for normal endpoints (no covariates) with known $\sigma^2_h$ and $\sigma^2_c$, using different power parameter determination methods. The posterior summary for each point on the curve is computed using $500$ simulation replicates. Note in this case we have $n=10$ but $m=40$.}
            \label{fig:13}
\end{figure}    
\begin{figure}
            \centerline{
            \includegraphics[scale=0.1]{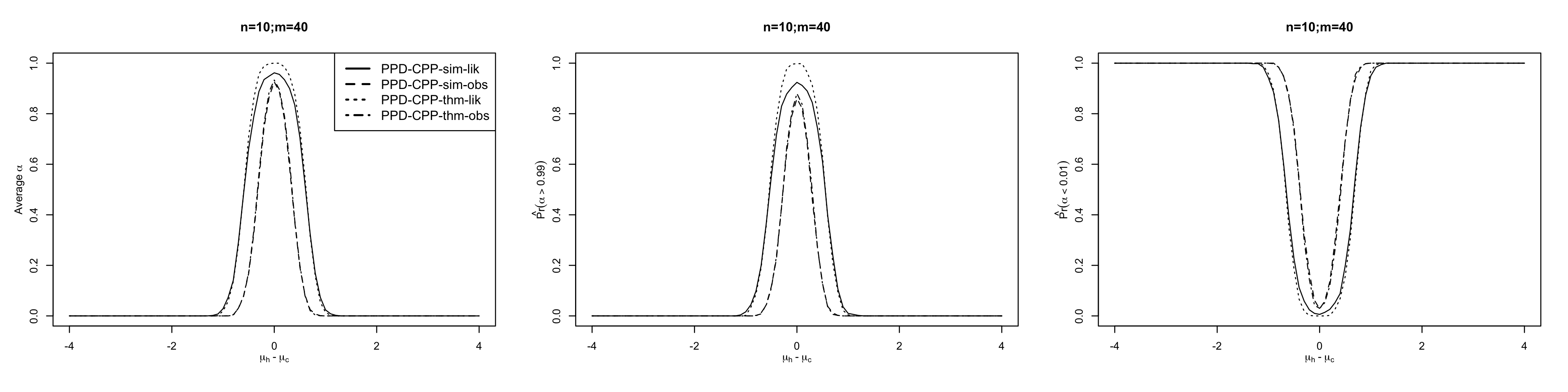}
            }
            \caption{Historical borrowing behavior comparison when $p_{CM}$ is derived either from Lemma \ref{lmma},\ref{lmma_obs} or computational method (\ref{pcm_compu}), with the change of mean difference for normal endpoints (no covariates), where $\sigma^2_h$ and $\sigma^2_c$ are assumed known.  The value on the curve is computed using $500$ power parameters. Note in this case we have $n=10$ but $m=40$.}
            \label{fig:14}
\end{figure}    
\begin{figure}
            \centerline{
            \includegraphics[scale=0.1]{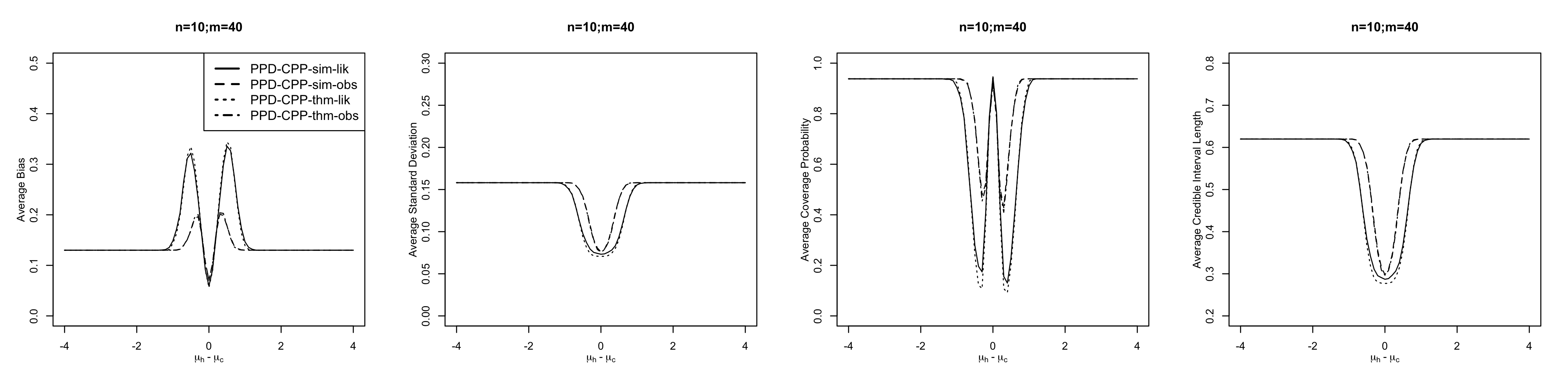}
            }
            \caption{Model performance comparison when $p_{CM}$ is derived either from Lemma \ref{lmma},\ref{lmma_obs} or computational method (\ref{pcm_compu}),for normal endpoints (no covariates) with known $\sigma^2_h$ and $\sigma^2_c$, using different power parameter determination methods. The posterior summary for each point on the curve is computed using $500$ simulation replicates. Note in this case we have $n=10$ but $m=40$.}
            \label{fig:15}
\end{figure}      

    \subsubsection*{Variance difference (known variance)}
    This simulation explores a scenario where data incongruence arises from differences in variance. We fix the means at $\mu_h = \mu_c = 20$, set the current variance $\sigma_c = 0.5$, and vary the historical variance $\sigma_h$ from $0.5$ to $1.5$ in increments of $0.1$. Both historical and current variances are assumed to be known. Based on the figure, the ``obs''-based PPD-CPP fails to detect the variance difference, resulting in a constant power parameter regardless of the level of incongruence. In contrast, the ``lik''-based PPD-CPP successfully captures the variance incongruence and demonstrates greater sensitivity than both EPP and CPP as the variance difference increases.

\begin{figure}
            \centerline{
            \includegraphics[scale=0.1]{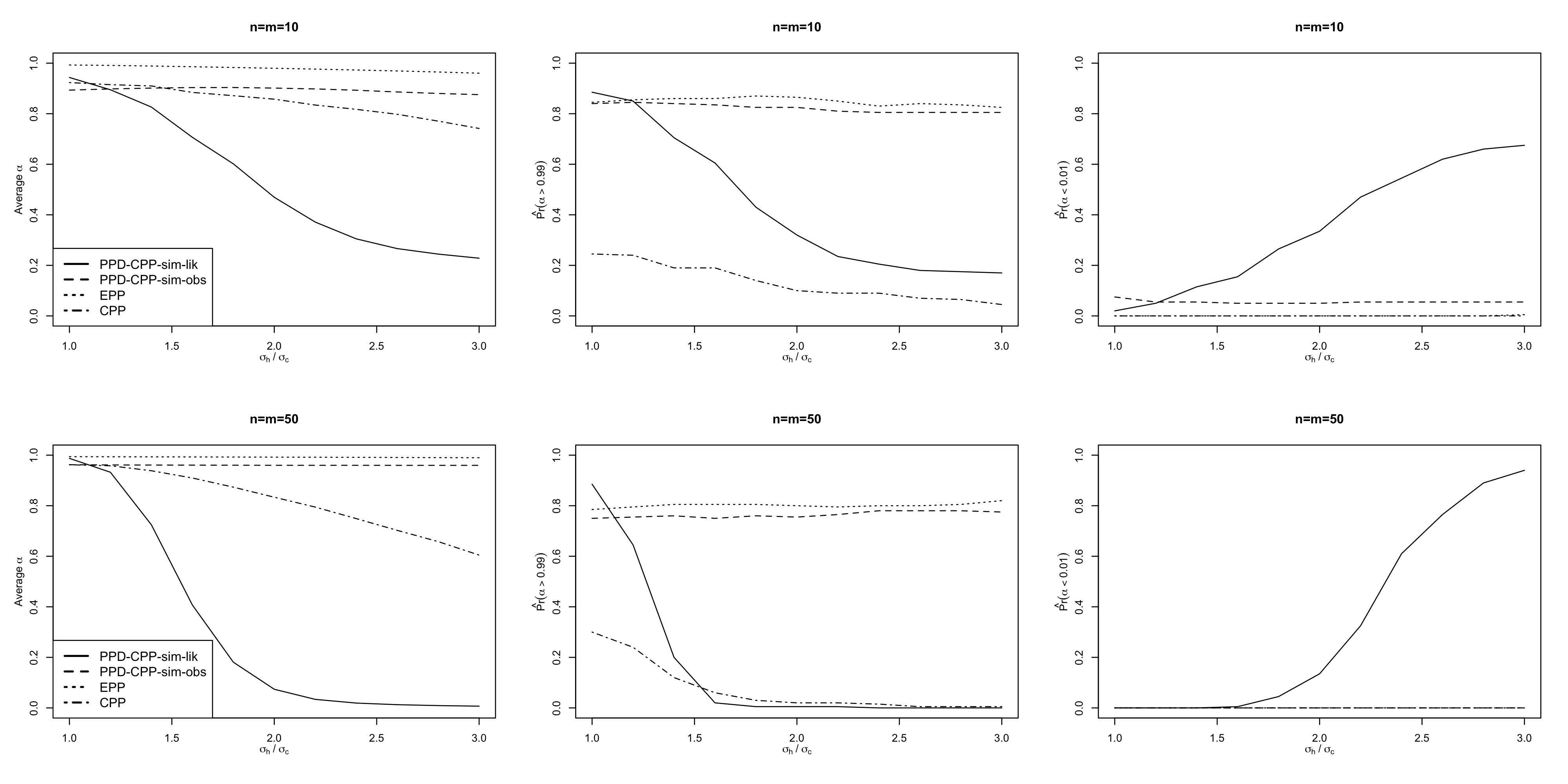}
            }
            \caption{Historical borrowing behavior with the change of variance difference for normal endpoints (no covariates), where $\mu_h$ and $\mu_c$ are assumed unknown. The value on the curve is computed using $500$ power parameters.}
            \label{fig:16}
\end{figure}
\begin{figure}
            \centerline{
            \includegraphics[scale=0.1]{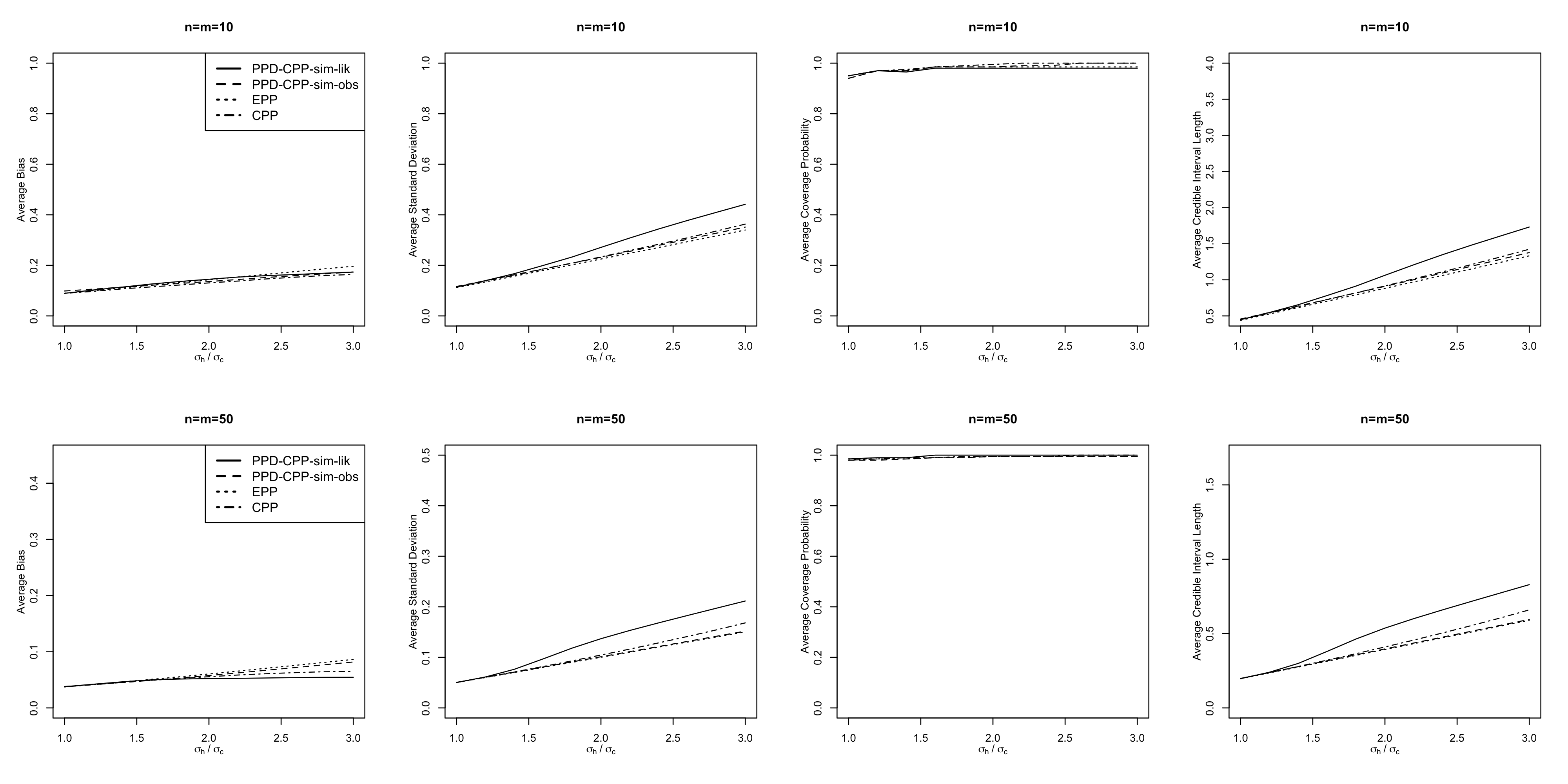}
            }
            \caption{Model performance for normal endpoints (no covariates) with known $\sigma^2_h$ and $\sigma^2_c$, using different power parameter determination methods. The incongruence between historical and current data are stemming from the variance difference. The posterior summary for each point on the curve is computed using $500$ simulation replicates.}
            \label{fig:17}
\end{figure}    
\begin{figure}
            \centerline{
            \includegraphics[scale=0.1]{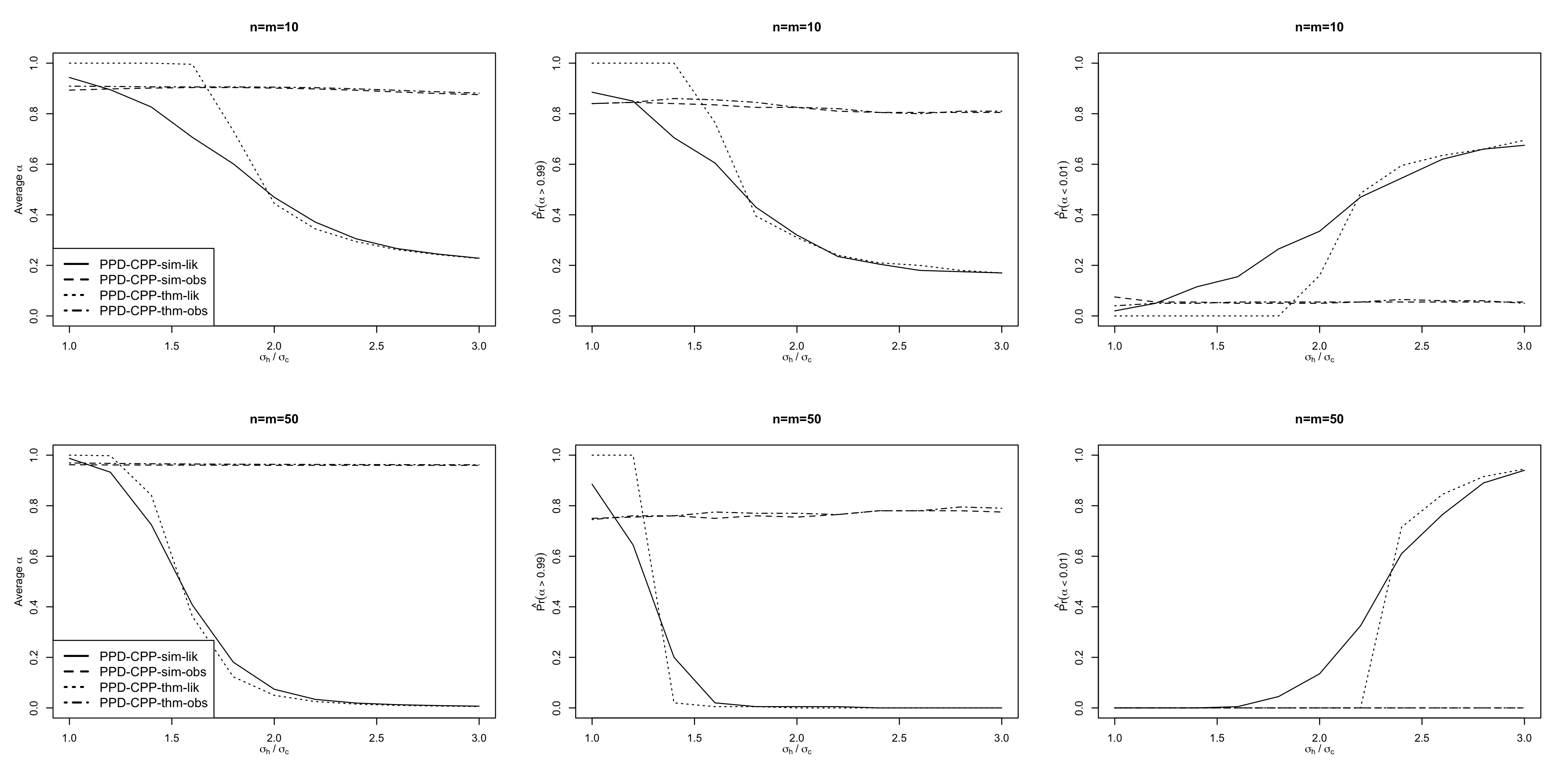}
            }
            \caption{Historical borrowing behavior comparison when $p_{CM}$ is derived either from Lemma \ref{lmma},\ref{lmma_obs} or computational method (\ref{pcm_compu}), with the change of variance difference for normal endpoints (no covariates), where $\sigma^2_h$ and $\sigma^2_c$ are assumed unknown.  The value on the curve is computed using $500$ power parameters.}
            \label{fig:18}
\end{figure}    
\begin{figure}
            \centerline{
            \includegraphics[scale=0.1]{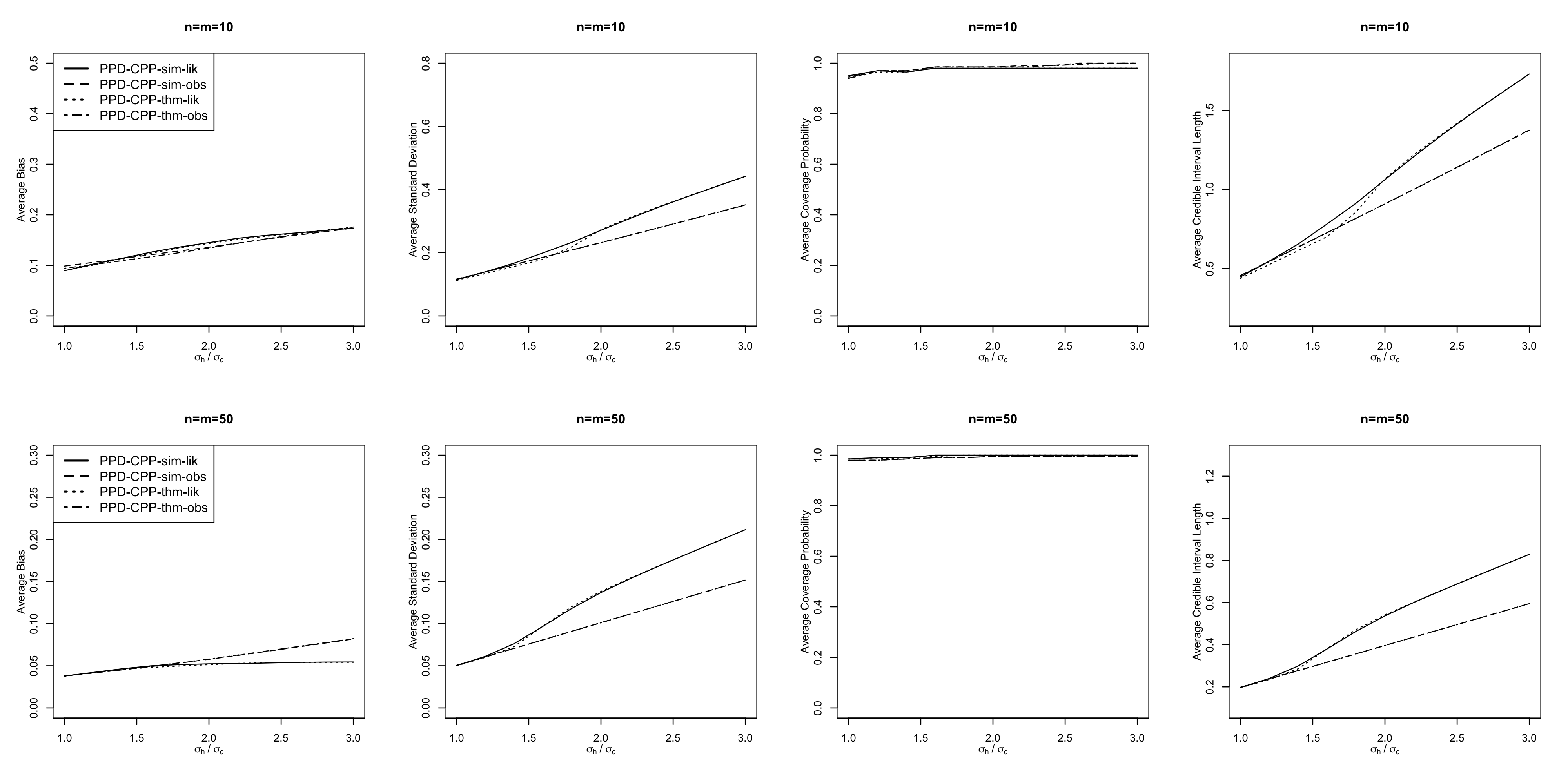}
            }
            \caption{Model performance comparison when $p_{CM}$ is derived either from Lemma \ref{lmma},\ref{lmma_obs} or computational method (\ref{pcm_compu}),for normal endpoints (no covariates) with unknown $\sigma^2_h$ and $\sigma^2_c$, using different power parameter determination methods. The incongruence between historical and current data are stemming from the variance difference. The posterior summary for each point on the curve is computed using $500$ simulation replicates. }
            \label{fig:19}
\end{figure}  
    
    \subsubsection*{Regression with covariate shifts}
\begin{figure}
            \centerline{
            \includegraphics[scale=1]{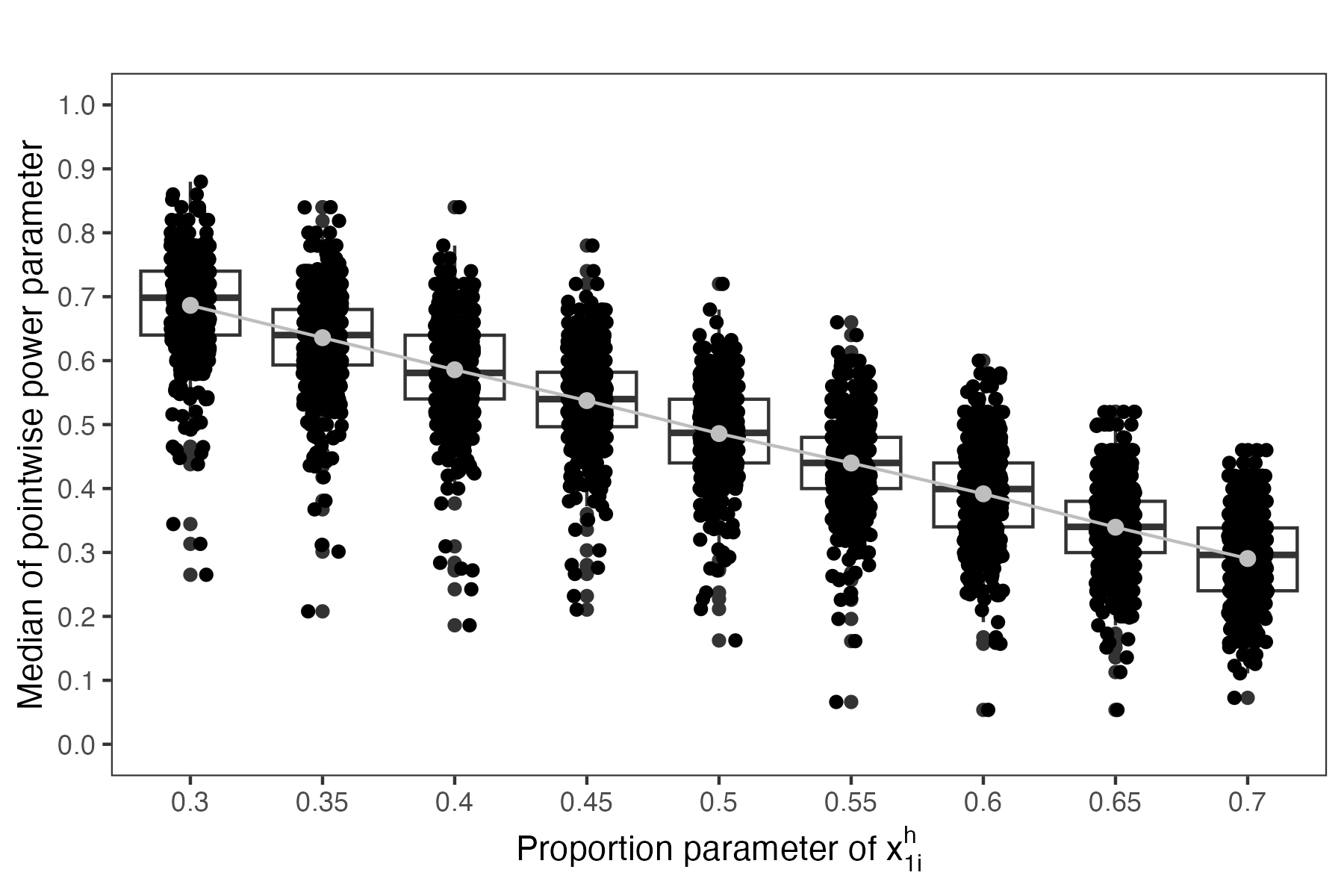}
            }
            \caption{The median of pointwise power parameter estimates VS the proportion parameter in the Bernoulli distribution used to simulate binary historical predictor $x_{1i}^h,i=1,\hdots,m$, from $500$ replicates of PPD-CPP-pw-lik. The grey dot represents the sample average. }
            \label{fig:20}
\end{figure}

    Following the simulation setup of the regression in the main context, we fix $x_{1i}^c\sim Bern(0.5)$ for current data but varying the proportion $p\in (0.3,0.7)$ by an increment of $0.05$ for historical binary covariate  $x_{1i}^h\sim Bern(p)$. As shown in Figure~\ref{fig:20},  the median of the pointwise power parameters centers around $1-p$, corresponding to the proportion of the covariate's reference group for historical data. This finding confirms the benefit of selectively borrowing compatile historical subgroup especially when the primary incongruence stems from the categorical covariate effects.
\end{spacing}

\end{document}